\documentclass[fleqn,usenatbib]{mnras}

\usepackage{newtxtext,newtxmath}

\usepackage[T1]{fontenc}
\usepackage{ae,aecompl}

\usepackage{graphicx}	
\usepackage{amsmath}	
\usepackage{amssymb}	

\usepackage{mathptmx}
\usepackage{txfonts}

\usepackage{pdflscape}
\usepackage{longtable}
\usepackage{supertabular}

\usepackage{pifont}
\newcommand{\cmark}{\ding{51}}%
\newcommand{\xmark}{\ding{55}}%

\usepackage{rotating}
\usepackage{footnote}
\usepackage{times}

\newcommand{\ha}{H$\alpha$}

\newcommand{\hii}{H~{\sc ii}}

\newcommand{\nitrogen}{[N\,{\sc ii}]}

\newcommand{\oxygeniii}{[O\,{\sc iii}]}

\newcommand{\oxygeni}{[O\,{\sc i}]}

\newcommand{\sulfurt}{[S\,{\sc ii}]}

\def\vhel{\ifmmode{V_{{\rm HEL}}}\else{$V_{{\rm HEL}}$}\fi}
\def\vsys{\ifmmode{V_{\rm sys}}\else{$V_{\rm sys}$}\fi}
\def\kms{\ifmmode{~{\rm km\,s}^{-1}}\else{~km~s$^{-1}$}\fi}
\def\vlsr{\ifmmode{v_{\rm lsr}}\else{$v_{\rm lsr}$}\fi}

\title[Compact planetary nebulae]{Compact planetary nebulae: Improved IR diagnostic criteria based on classification tree modelling}

\author[Stavros Akras]{
Stavros Akras$^{1}$\thanks{E-mail: stavrosakras@gmail.com},
Lizette Guzman-Ramirez$^{2}$, Denise R. Gon\c{c}alves$^{3}$
\\
$^{1}$Instituto de Matem\'{a}tica, Estat\'{i}stica e F\'{i}sica, Universidade Federal do Rio Grande, Rio Grande 96203-900, Brazil\\
$^{2}$Leiden Observatory, Leiden University, Niels Bohrweg 2, 2333 CA Leiden, Netherlands\\
$^3$Observat\'orio do Valongo, Universidade Federal do Rio de Janeiro, Ladeira Pedro Antonio 43, 20080-090, Rio de Janeiro, Brazil\\
}

\date{Accepted XXX. Received YYY; in original form ZZZ}

\pubyear{2019}

\begin{document}
\label{firstpage}
\pagerange{\pageref{firstpage}--\pageref{lastpage}}
\maketitle

\begin{abstract}
    Planetary nebulae (PNe) are strong \ha\ line-emitters and a lot of new PNe discoveries have been made by the SuperCOSMOS AAO/UKST \ha\ Survey (SHS) and the Isaac Newton Telescope Photometric \ha\ Survey (IPHAS). However, the list of auto-generated \ha-excess candidates from these surveys as well as any photometric survey, prior to spectroscopic follow-up to confirm their nature, contains all varieties of \ha-line emitters like young stellar objects (YSOs), \hii~regions, compact PNe and emission line stars of all kinds. The aim of this work is to find new infrared criteria that can better distinguish compact PNe from their mimics using a machine learning approach and the photometric data from the 2MASS and WISE surveys. Three classification tree models have been developed with the following colour criteria: {\it W1-W4}$\geq$7.87 and {\it J-H}$<$1.10; {\it H-W2}$\geq$2.24 and {\it J-H}$<$0.50; and {\it Ks-W3}$\geq$6.42 and {\it J-H}$<$1.31 providing a list of candidates, characterised by a high probability to be genuine PNe. The contamination of this list of candidates from \ha\ mimics is low but not negligible. By applying these criteria to the IPHAS list of PN candidates and the entire IPHAS and VPHAS+ DR2 catalogues, we find 141 sources, from which 92 are known PNe, 39 are new very likely compact PNe (without an available classification or uncertain) and 10 are classified as \hii~regions, Wolf-Rayet stars, AeBe stars and YSOs. The occurrence of false positive identifications in this technique is between 10 and 15~per cent.
\end{abstract}




\section{Introduction}

Over the past twenty years, a significant effort has been devoted to discover new Planetary Nebulae (PNe) doubling their Galactic population from $\sim$1600 \citep{Acker1992} to $\sim$3500 \citep{Parker2016}. The main reason for the rapid increase of these new discoveries is the wide-field, arc-second spatial resolution, narrow-band \ha\ photometric surveys, such as the SuperCOSMOS AAO/UKST \ha\ Survey (SHS) of the southern Galactic plane \citep[]{Parker2005,Frew2014} and the Isaac Newton Telescope Photometric \ha\ Survey (IPHAS) of the northern Galactic plane \cite[]{Drew2005}. Both surveys have unveiled numerous \ha\ point-sources including PNe, \hii~regions, symbiotic stars (SySts), Be stars, Wolf-Rayet stars and young stellar objects (YSO), among others \citep[see][]{Parker2006,Miszalski2008,Witham2008,Corradi2008,Viironen2009a,Viironen2009b,Sabin2014}. Yet, the final identification of these \ha\ sources was made only after follow-up spectroscopy.

\citet{Parker2006} presented the first release of the Macquarie/AAO/Strasbourg \ha\ 
PNe catalogue (MASH) of over 900 true (spectroscopically confirmed) and candidate PNe (classified as likely or possible) found in the SHS survey. These numbers corresponded to  almost 60 per cent increase of the Galactic PNe population. A couple of years later, \citet{Miszalski2008} increased the number of known and candidate PNe in the MASH catalogue by detecting around 300 more members in the fields of SHS survey. All these discoveries have made the SHS survey one of the most important in the field of PNe and \ha-emitters in general.

Besides the SHS survey, several PNe have also been discovered in the IPHAS survey either, by identifying candidates in the {\it r$^\prime$}-\ha\ vs. {\it r$^\prime$}-{\it i$^\prime$} diagnostic colour-colour diagrams \citep[hereafter DCCD,][]{Viironen2009a,Viironen2009b} or by visually inspecting the IPHAS fields \citep[]{Sabin2014}. Different techniques resulted in different groups of PNe. In particular, the visual inspection of the IPHAS fields revealed extended, low-surface brightness PNe while the IPHAS DCCD brought out compact and young PNe. A list of 781 candidate and 224 known PNe in the IPHAS survey was published in \citet{Viironen2009b}. Although, IPHAS data are still not fully explored and four new PNe were found and spectroscopically confirmed later on by \citet{Hsia2014}.

Besides the discoveries from these two \ha\ surveys, more Galactic PNe have been recently found in the CORNISH survey (90, \cite{Irabor2018}; 62, \cite{Fragkou2018}), and in the UWISH2 survey (183, \cite{Gledhill2018}). Smaller private surveys have also contributed to the discovery of new Galactic PNe \citep[e.g.][]{Boumis2003,Boumis2006}. 

The optical ({\it \ha--R} vs. {\it R--I} or {\it r$^\prime$--\ha} vs. {\it r$^\prime$-i$^\prime$}) and the 2MASS ({\it J--H} vs. {\it H--Ks}) DCCDs have been extensively used to seek for PNe or SySt candidates \citep[e.g.][]{Miszalski2008,Corradi2008,Viironen2009b,Corradi2010,Rodriguez2014,Akras2019}. The first two DCCDs in the optical regime assure that the sources are in fact strong \ha-line emitters, while the third DCCD in the near-infrared regime guarantees that they are also dusty sources. However, follow-up spectroscopic surveys have shown that both lists of candidate PNe and SySts derived from the IPHAS catalogue are heavily contaminated from other types of \ha-emitters like Be stars, YSOs and \hii~regions.

The cause of this high contamination is the lack of additional information from longer wavelengths. \citet{Akras2019} have shown that information from the WISE survey \citep[]{Wright2010} can significantly improve the identification of SySts resulting in less contaminated lists/catalogues. \citet{Cohen2007} performed a multi-wavelength study on the PNe found in the MASH catalogue using the photometric data from the Infrared Array Camera (IRAC), on the Spitzer telescope, and proposed a specific locus for PNe on the [3.6]--[4.5] vs. [5.8]--[8.0] DCCD. This locus better distinguishes PNe from other \ha-emitters but not from YSOs. 
 
Motivated by the recent work of \citet{Akras2019} and its success in identifying new SySt candidates using a new set of criteria extracted from a machine learning approach, we applied the same methodology and sought for those infrared criteria that better discriminate PNe from other \ha-sources. Then, by applying these new criteria to the list of PN candidates in the IPHAS catalogue \citep[]{Viironen2009b}, we provide a new, less contaminated from \ha\ mimics, list of compact IPHAS PN candidates. The same criteria are also applied to the entire VPHAS+ DR2 catalogue \citep[]{Drew2016}, on which no one has yet searched for PNe.

The paper is organized as follow: the classification tree models are presented and discussed in Section~2. In Section~3, the distribution of known PNe as a function of their angular radius is explored. In Section~4, our criteria are applied to the list of candidate PNe, from \citet{Viironen2009b}, and the VPHAS+ DR2 catalogue. New PN candidates as well as known PNe that satisfy the criteria of our classification tree models are presented. In addition, we also apply our criteria to the entire IPHAS DR2 catalogue \citep{Barentsen2014}, in order to verify the population of known or PN candidates in this survey. Our conclusions are summarized in Section~5. 

\section{Classification tree algorithm} 

According to the DCCDs presented by Akras and collaborators, PNe and YSOs occupy different regions in most of the diagrams \citep{Akras2019}. Instead of exploring  all the possible combinations of 2MASS and WISE DCCDs that may be good to identify new PNe, a machine learning approach, the classification tree algorithm, is used.

The {\it evtree} function \citep{Grubinger2014} (Grubinger, Zeileis \& Pfeiffer 2014) in R software \citep{Rcoreteam},  as well as a set of 10 colour indices ({\it J--H}, {\it H--K$_s$}, {\it K$_s$--W1}, {\it W1--W2}, {\it W2--W3}, {\it W3--W4}, {\it J--W1}, {\it H--W2}, {\it K$_s$--W3}, and {\it W1--W4}) were considered to built our classification tree models. Despite the fact that these 10 colour indices cover only a part of possible combinations between the 2MASS and WISE bands, they consist of a representative set for the near and mid-IR regime.

To train our models, we made use of the lists of \ha\ emitters in \citet{Akras2019}. This list contains several types of sources that mimic PNe, either in the optical or infrared regime:  220 SySts \citep{Akras2019b}, 162 Wolf-Rayet stars \citep{derHucht2001}, 185 Be \citep{chojnowski2015}, 173 AeBe stars \citep{Rodrigues2009,Vieira2003,Herbst1999}, 191 cataclysmic variables \citep{Hoard2002}, 183 classical \citep{Galli2015,France2014,Grankin2007,Herbst1999} and 213 weak T Tauri stars \citep{Galli2015,Grankin2008,Cieza2007,Herbst1999}, 316 Mira stars \citep{Huemmerich2012,Whitelock2008}, and 260 YSOs \citep{Rebull2011,Harvey2007}. All these different type of sources were merged in our category which called "Mimics". The classification tree algorithm was used to find the models, or the set of criteria, that provide an adequate partition between PNe and Mimics.

\begin{figure*}
\vbox{
\includegraphics[scale=0.7]{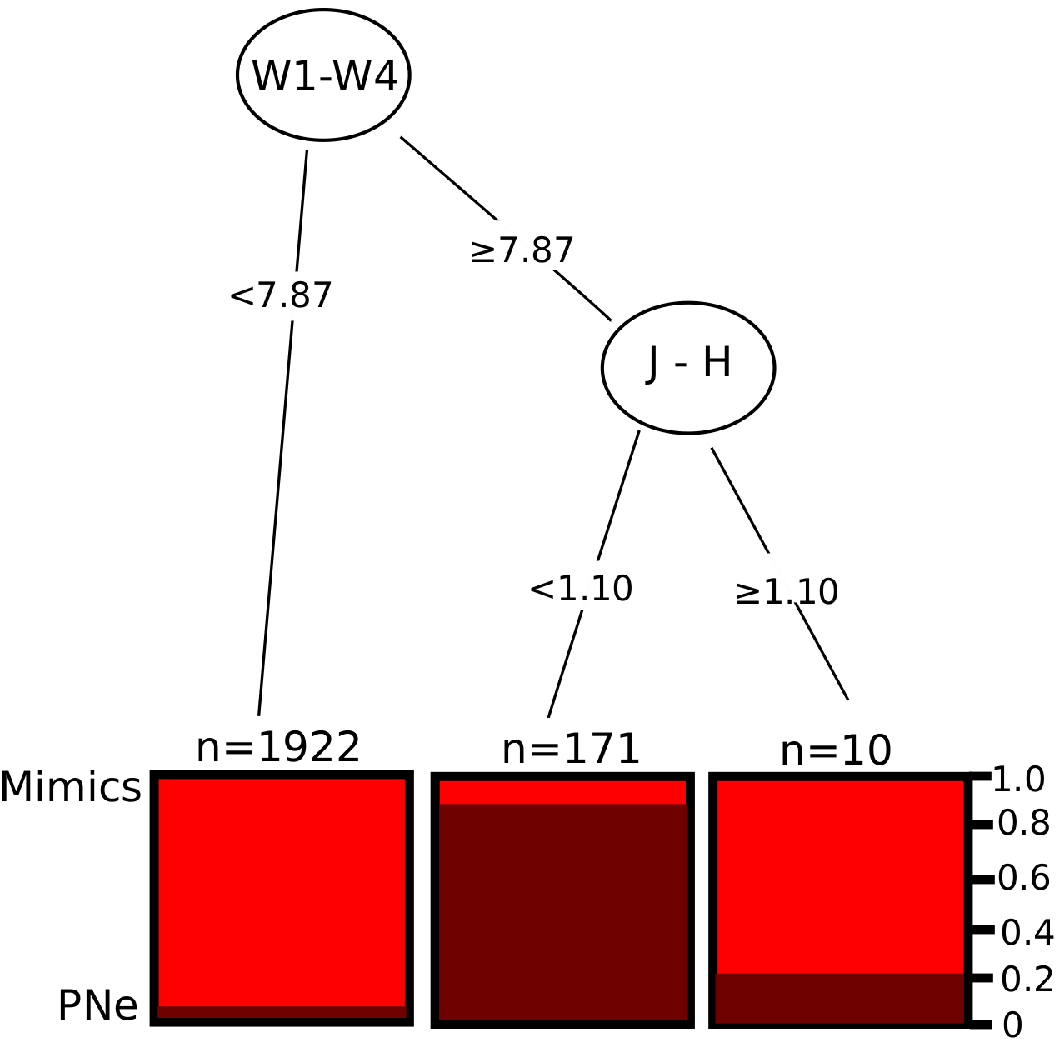}
\includegraphics[scale=0.7]{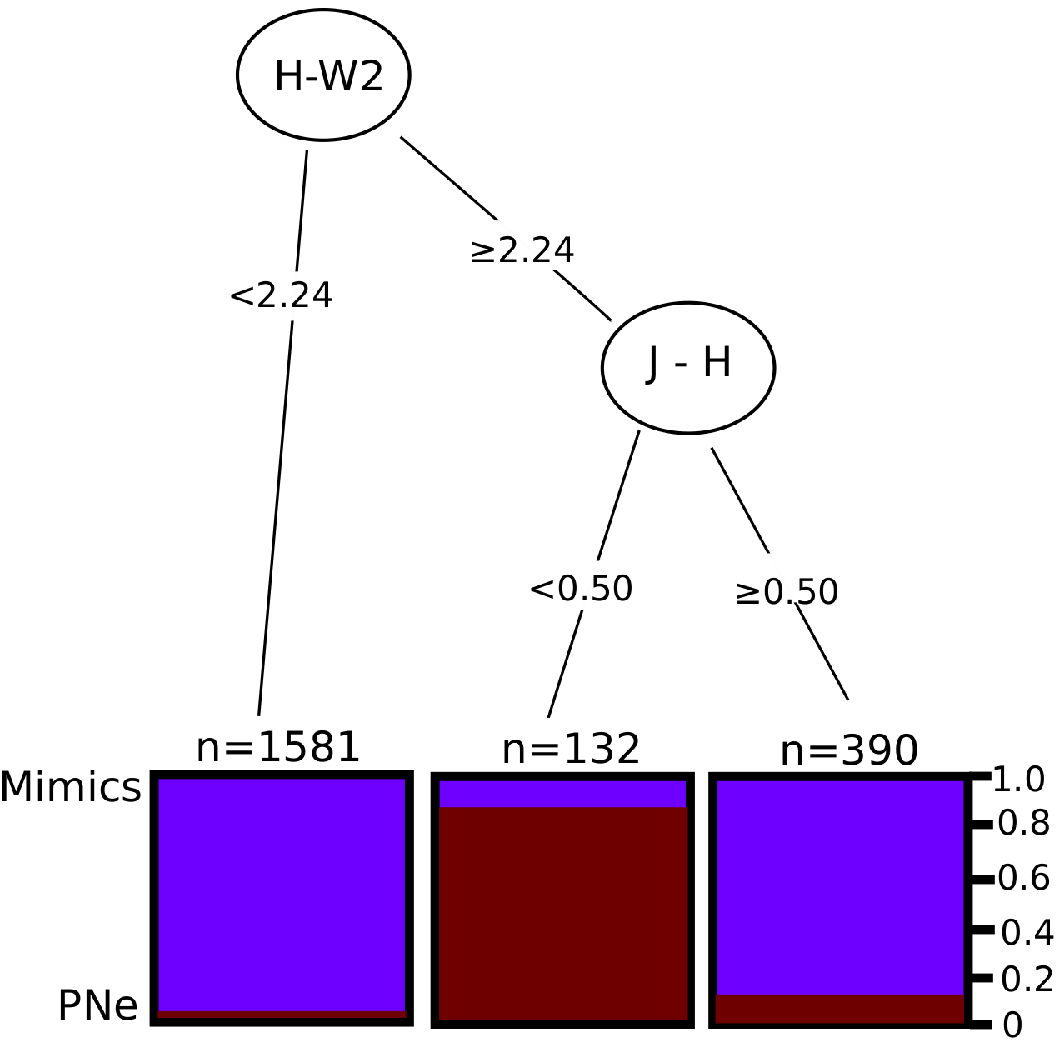}
\includegraphics[scale=0.7]{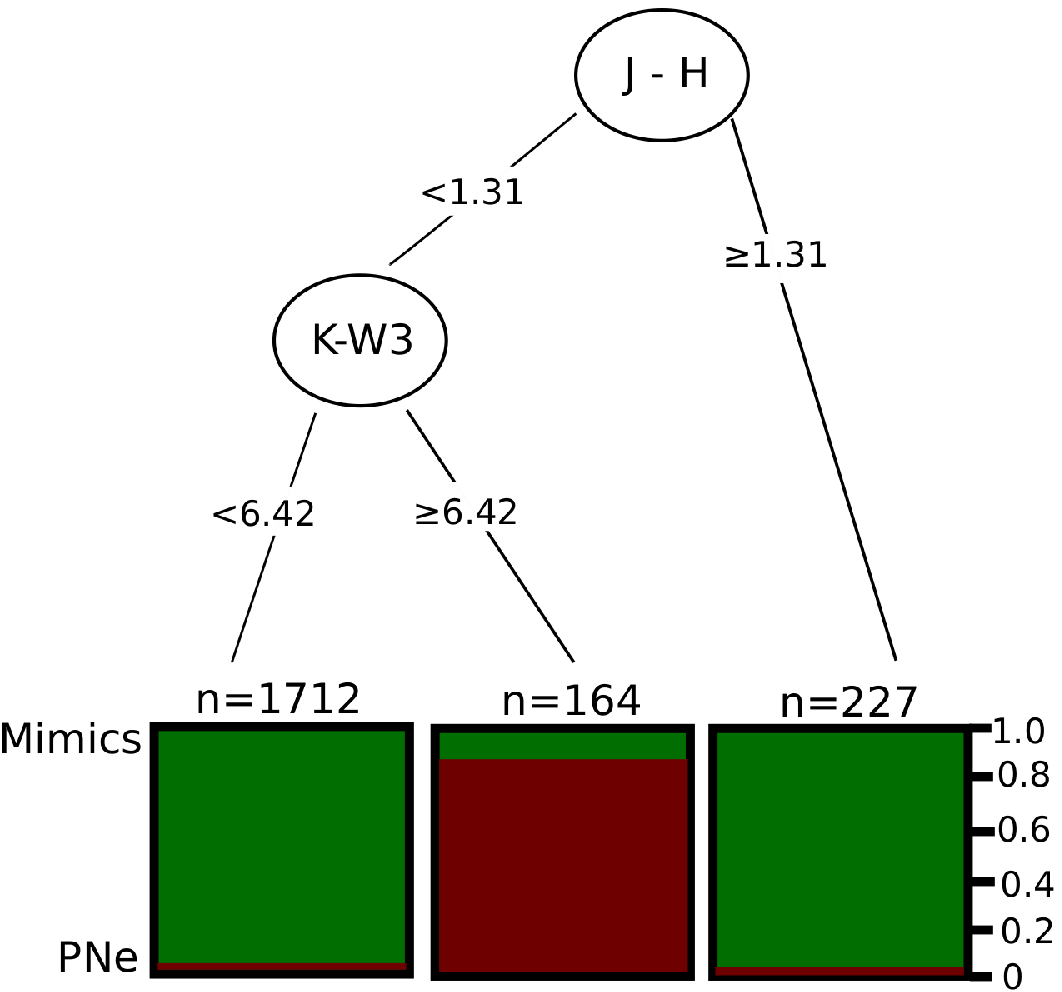}}
\caption[]{Classification tree models: top-left panel, Model~1 (M1); top-right panel, Model~2 (M2) and lower panel Model~3 (M3).}
\label{fig1} 
\end{figure*}

Three different classification tree models are found and their tree plots are presented in Figure \ref{fig1}. All three models show that two criteria are enough to distinguish PNe: {\it W1-W4}$\geq$7.87 and {\it J-H}$<$1.10 (hereafter model M1), {\it H--W2}$\geq$2.24 and {\it J-H}$<$0.503 (model M2) and {\it J-H}$<$1.31 and {\it Ks--W3}$\geq$6.42 (model M3).

Model M1 correctly classifies 171 out of 188 PNe (91 per cent) with only 6 per cent of contamination from other classes of \ha-emitters. Model M2 identifies 132 PNe or 70 per cent and it shows a slightly higher contamination from mimics (9 per cent). Finally, the model M3 classifies 164 PNe or 87 per cent of the training list with a 8 per cent of contamination from mimics. The overall occurrence of false positive identifications is estimated between 10 and 15 per cent comparable to those of SySts in \citet{Akras2019}. For this task, we repeated the algorithm a few time using as training and testing sets, randomly selected samples of 80 and 20 per cent of the PNe sample, respectively. We argue that all the three models are suitable for searching/identifying new Galactic PNe providing at the same time a less contaminated list of candidates, with respect to the usual optical and 2MASS DCCDs.

Because YSOs and compact \hii~regions are also very dusty sources and may exhibit {\it W1--W4}, {\it H--W2} and {\it K--W3} colour indices similar to PNe, additional lists of these two sources were used to further verify the contamination level of the resulting list derived from the M1, M2 and M3 criteria.  

Fifty new YSOs were discovered in the direction of the Cygnus OB2 region through their IPHAS and 2MASS counterparts. Follow-up spectroscopy revealed their nature as T~Tauri and Herbig Ae stars \citep[][]{Vink2008}. We gathered the AllWISE photometric data of the Vink's YSOs and calculated their {\it W1--W4}, {\it H--W2} and {\it Ks-W3} colour indices in order to compare them with those from the training samples of T~Tauri stars and YSOs. Only 8~per cent of the YSOs in Vink's list satisfy all the criteria, while 20~per cent of them pass only the criteria of the {\it W1--W4}, {\it H--W2} and {\it Ks-W3} colour indices.

In order to further explore the distributions of YSOs, compact \hii~regions and PNe in the colour space derived from the classification tree models, we used the Red MSX Source (RMS) survey \citep[][]{Lumsden2013}, a multi-wavelength survey that provides genuine YSOs well separated from PNe and \hii~regions. Cross-matching the RMS and AllWISE catalogues, we found 139 matches either with 2MASS or AllWISE photometric data. Only 53 of them have both counterparts and are classified as follow: 22 as \hii~regions, 23 as PNe and 8 as YSOs \citep[][]{Urquhart2009}.

\begin{figure*}
\vbox{
\includegraphics[width=\columnwidth]{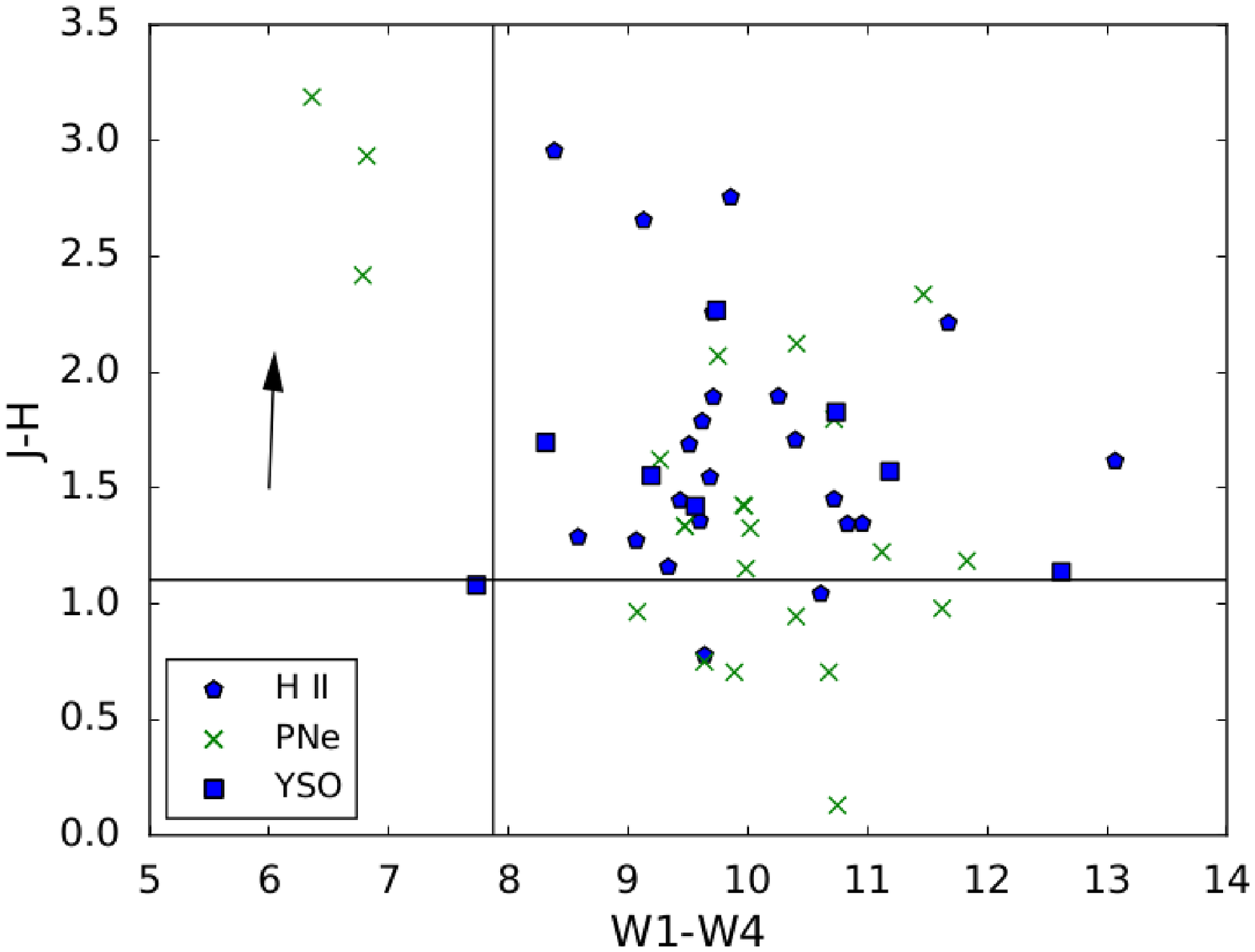}
\includegraphics[width=\columnwidth]{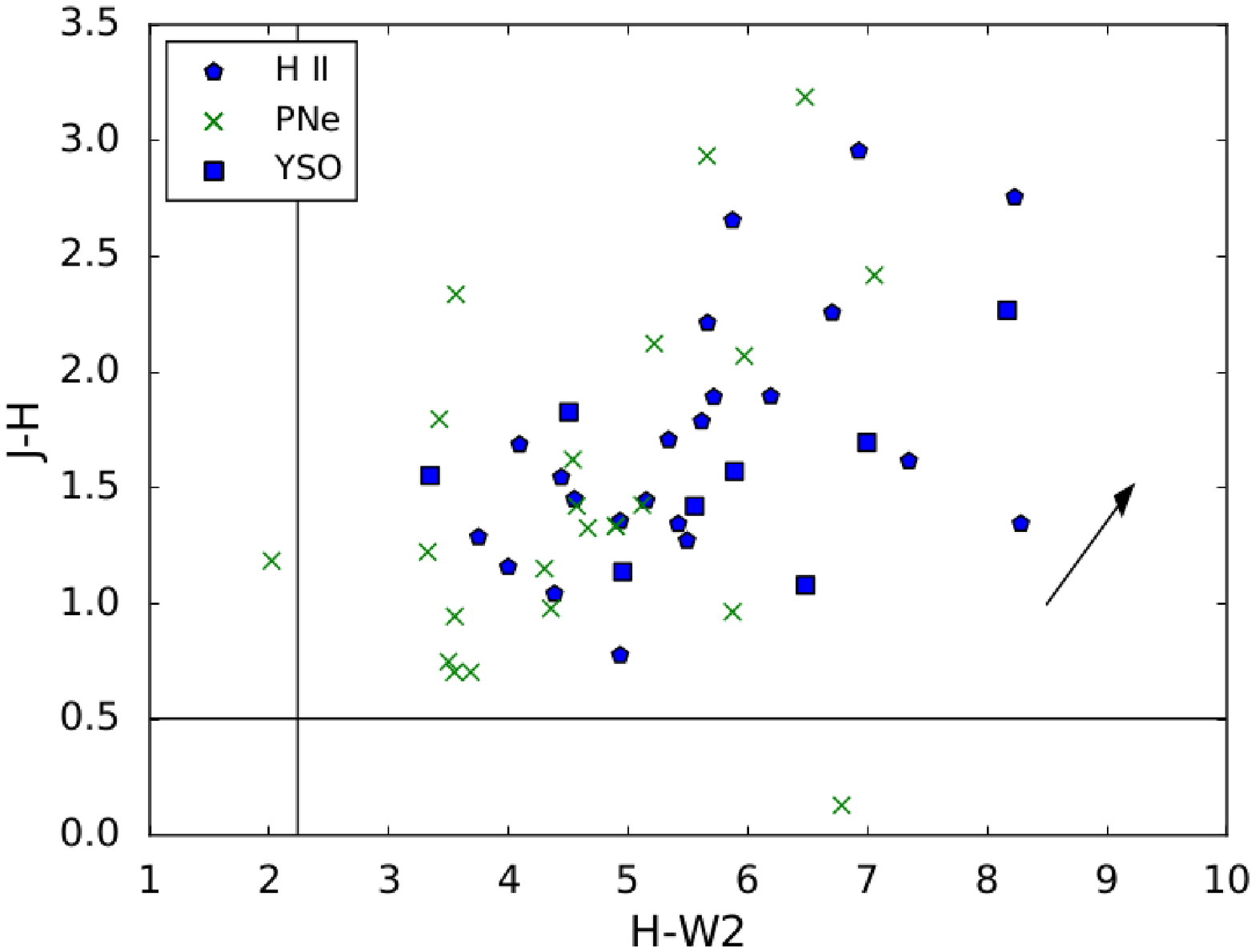}
\includegraphics[width=\columnwidth]{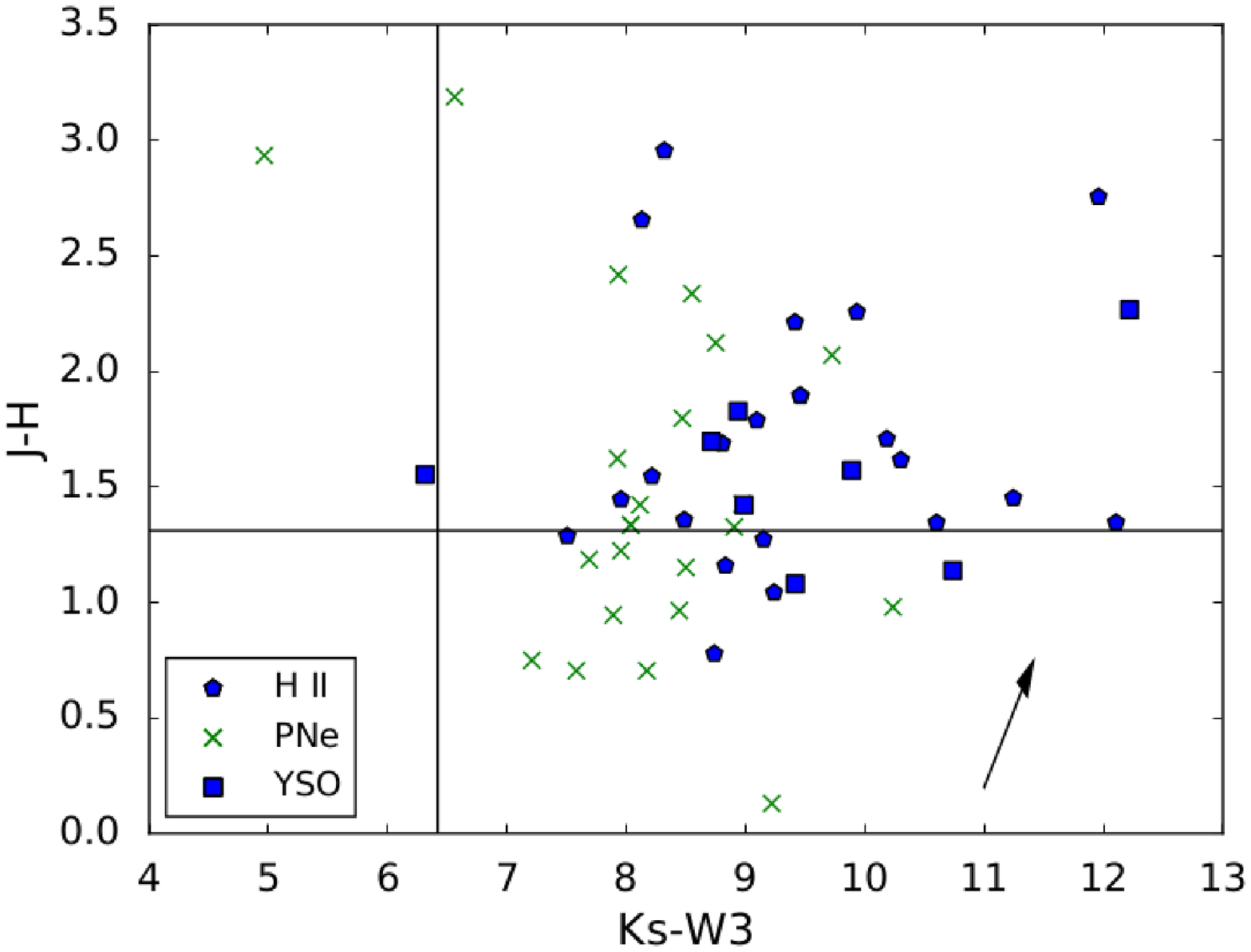}}
\caption[]{The distribution of PNe (green "x"-symbols), \hii~regions (blue pentagons), YSOs (blue squares) and possible YSOs or \hii~regions (blue triangles) in the RMS survey, for the colour indices criteria derived by the classification tree algorithm: M1 (upper-left panel), M2 (upper -right panel) and M3 (lower panel). The vertical and horizontal lines illustrate the criteria of each model (see Fig.\ref{fig1}). The black arrows correspond to 4~mag extinction in the V band.}
\label{fig2} 
\end{figure*}

By studying the distribution of these PNe, YSOs and \hii~regions in the colour spaces of our classification tree models (Fig. \ref{fig2}) -- {\it W1--W4}/{\it J--H}, {\it H--W2}/{\it J--H} and {\it Ks--W3}/{\it J--H} spaces -- we conclude that: 
\begin{itemize}
\item All YSOs (except one), \hii~regions and PNe (except three) exhibit {\it W1--W4}$\geq$7.87 and they can have {\it J--H} colour index either lower or higher than 1.10. Seven out of 22 PNe (32 per cent) are found to satisfy both criteria from the M1 model, contaminated only by two non-PNe sources. The M1 model likely provides a significantly less contaminated list of candidate PNe than in previous attempts. However, there is also a population of PNe with {\it J--H}$>$1.10 which is heavily contaminated by mimics and very difficult to be identified.
\item All YSOs, \hii~regions and PNe (except from one) display {\it H--W2}$\geq$2.24 but none of them satisfy the second criterion of M2 model ({\it J--H}$<$0.503). We thus argue that any source that satisfies M2 criteria is a candidate, characterised by a high probability to be genuine PNe.
\item All YSOs (except one), \hii~regions and PNe (except one) are found to have {\it Ks--W3}$\geq$6.42 and they are divided into two subgroups according to the {\it J--H} colour. The M3 criteria correctly identify 11 out of 22 PNe (50 per cent). Seven non-PNe sources also exhibit {\it J--H}$<$1.31. Similar to the M1 model, PNe can also exhibit {\it J--H}$>$1.31, but they are heavily mixed with non-PNe sources.
\end{itemize}

Overall, if a source satisfies the criteria of all the tree models it is a highly probable genuine PN. Good PN candidates can also be obtained from the criteria of the M1 and M3 models. Figure~2 illustrates the {\it W1--W4} vs. {\it J--H}, {\it H--W2} vs. {\it J--H} and {\it Ks--W3} vs. {\it J--H} DCCDs for the 53 RMS sources.

\section{Compact PNe}
\begin{figure*}
\includegraphics[width=\columnwidth]{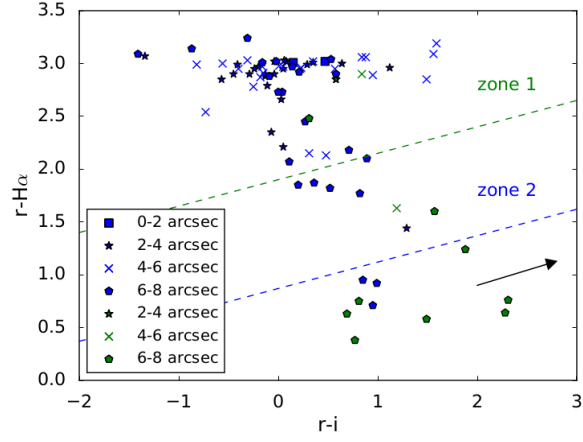}
\caption[]{r$^\prime$-\ha\ vs. r$^\prime$-i$^\prime$ DCCD for PNe of different angular radius: 0-2~arcesec (square); 2-4~arcsec (star); 4-6~arcsec (cross) and 6-8~arcsec (pentagon). The blue and green symbols correspond to the IPHAS and VPHAS+ PNe, respectively. Zones 1 and 2 defined by \citet{Viironen2009b} are also indicated by the green and blue dashed-lines. The black arrow indicates to 3~mag extinction in the V band.}
\label{fig3}
\end{figure*}

Due to the fact that we are interested in finding compact PNe hidden in photometric catalogues, we also explored the distribution of known PNe for different angular radius: 0-2, 2-4, 4-6 and 6-8~arcsec. For this exercise, the HASH catalogue \citep[]{Parker2016} was cross-matched with the IPHAS and the VPHAS+ catalogues resulting in 71 and 13 matches, respectively. Their position in the IPHAS ({\it r$^\prime$}-\ha) vs. ({\it r$^\prime$-i$^\prime$}) DCCD is presented in Figure~\ref{fig3}. 

Most PNe (66 out of 84 or 79 per cent) are found to lie in zone~1 ({\it r$^\prime$--\ha} $>$ 0.25({\it r$^\prime$--i$^\prime$}) + 1.9) rather than in zone~2 (8 PNe or 9 per cent, 0.25({\it r$^\prime$--i$^\prime$}) + 0.87) $<$ {\it r$^\prime$--\ha} $<$ 0.25({\it r$^\prime$--i$^\prime$}) + 1.9) \citep{Viironen2009b}, while 10 of them or 12 per cent do not satisfy the IPHAS criteria at all. From the eight PNe in the zone~2, almost all of them (one has angular radius in the range of 2-4~arcsec and one in the range of 4-6~arcsec) are relatively large with radius higher than 6~arcsec. As for those below zone~2, all have angular radius larger than 6~arcsec. This is consistent with their low ({\it r$^\prime$}-\ha) colour index compared to the more compact ones. Generally, the distribution of the compact HASH PNe in the IPHAS DCCD is consistent with previous studies \citep[e.g.][]{Viironen2009a,Viironen2009b}.

\begin{figure*}
\vbox{
\includegraphics[width=\columnwidth]{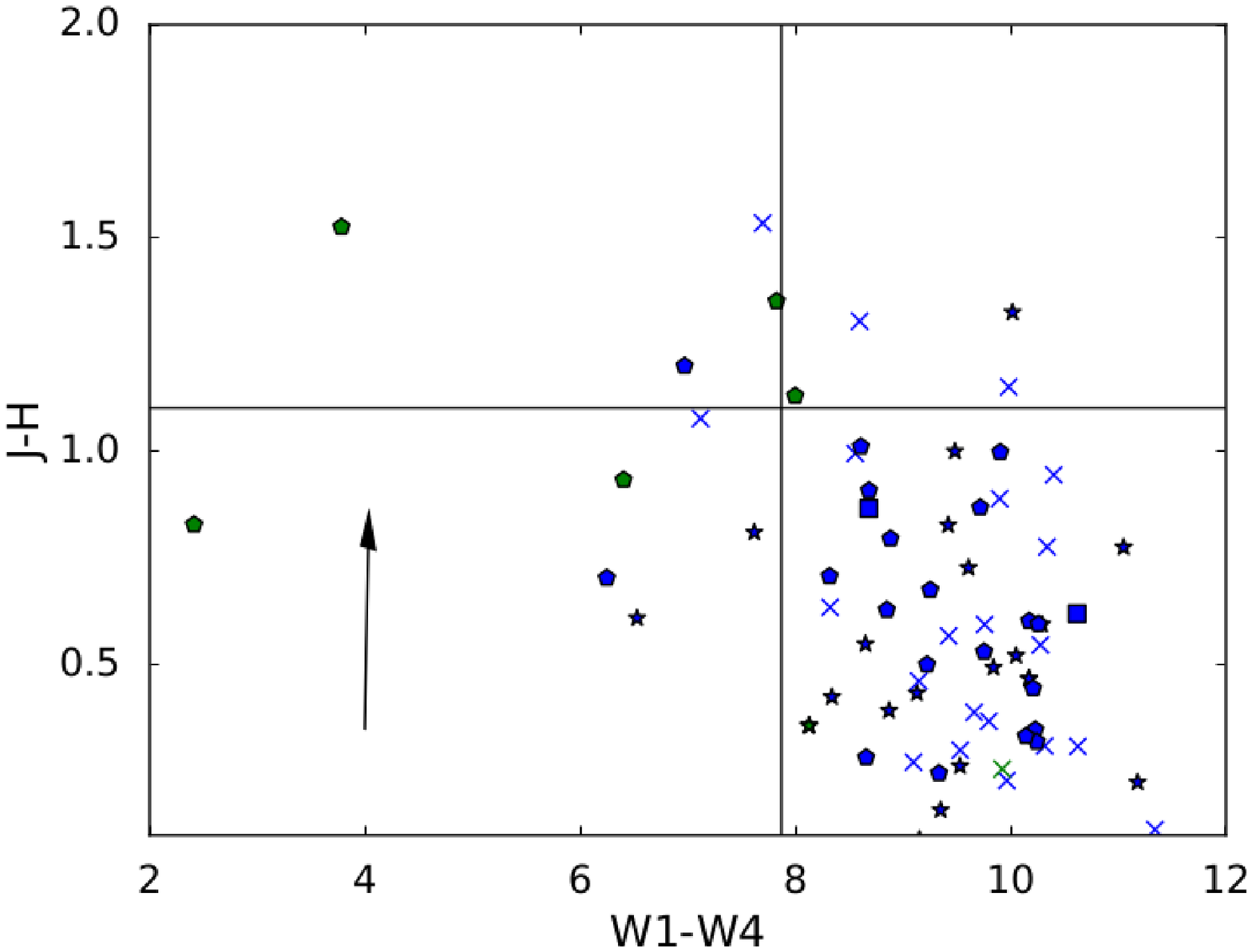}
\includegraphics[width=\columnwidth]{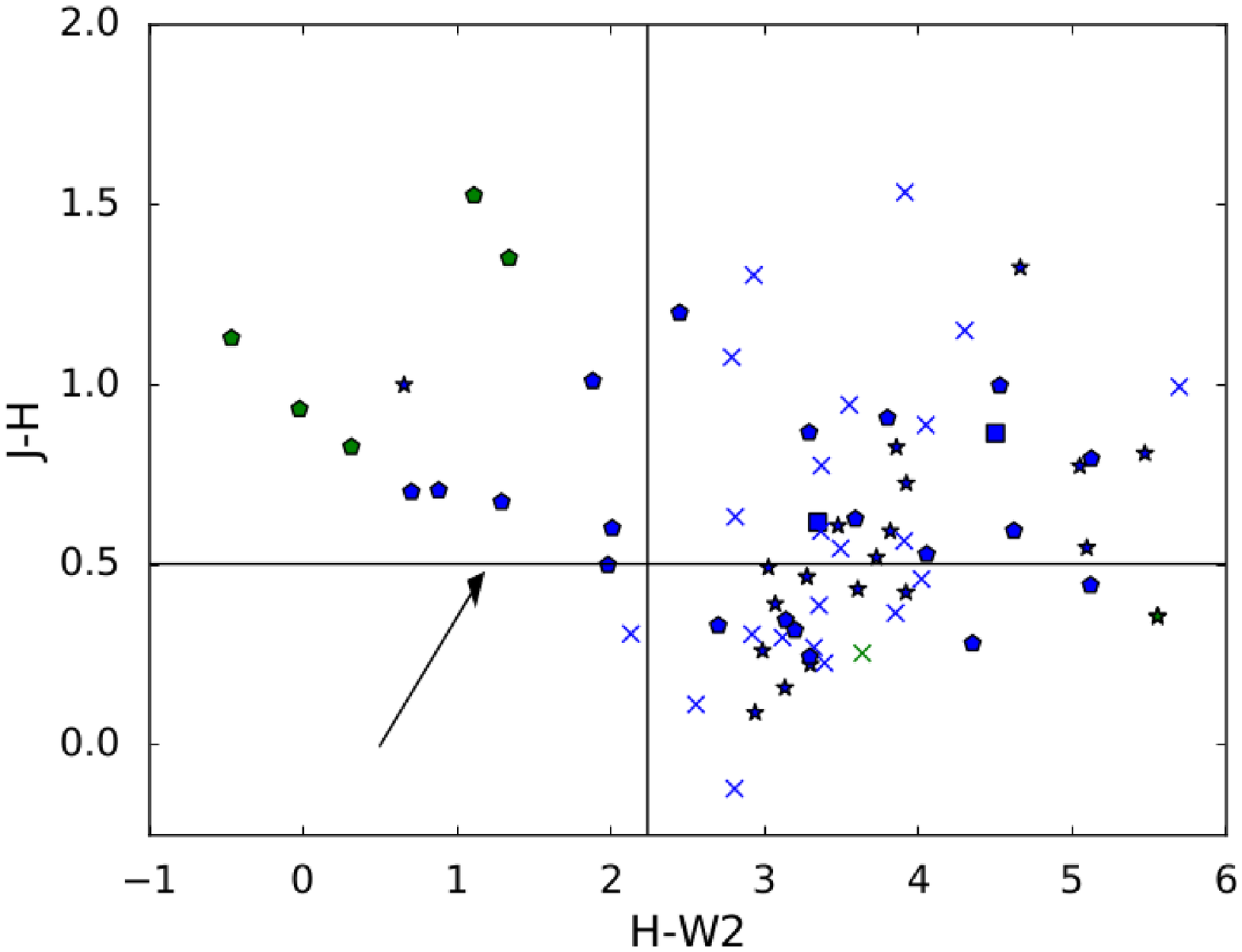}
\includegraphics[width=\columnwidth]{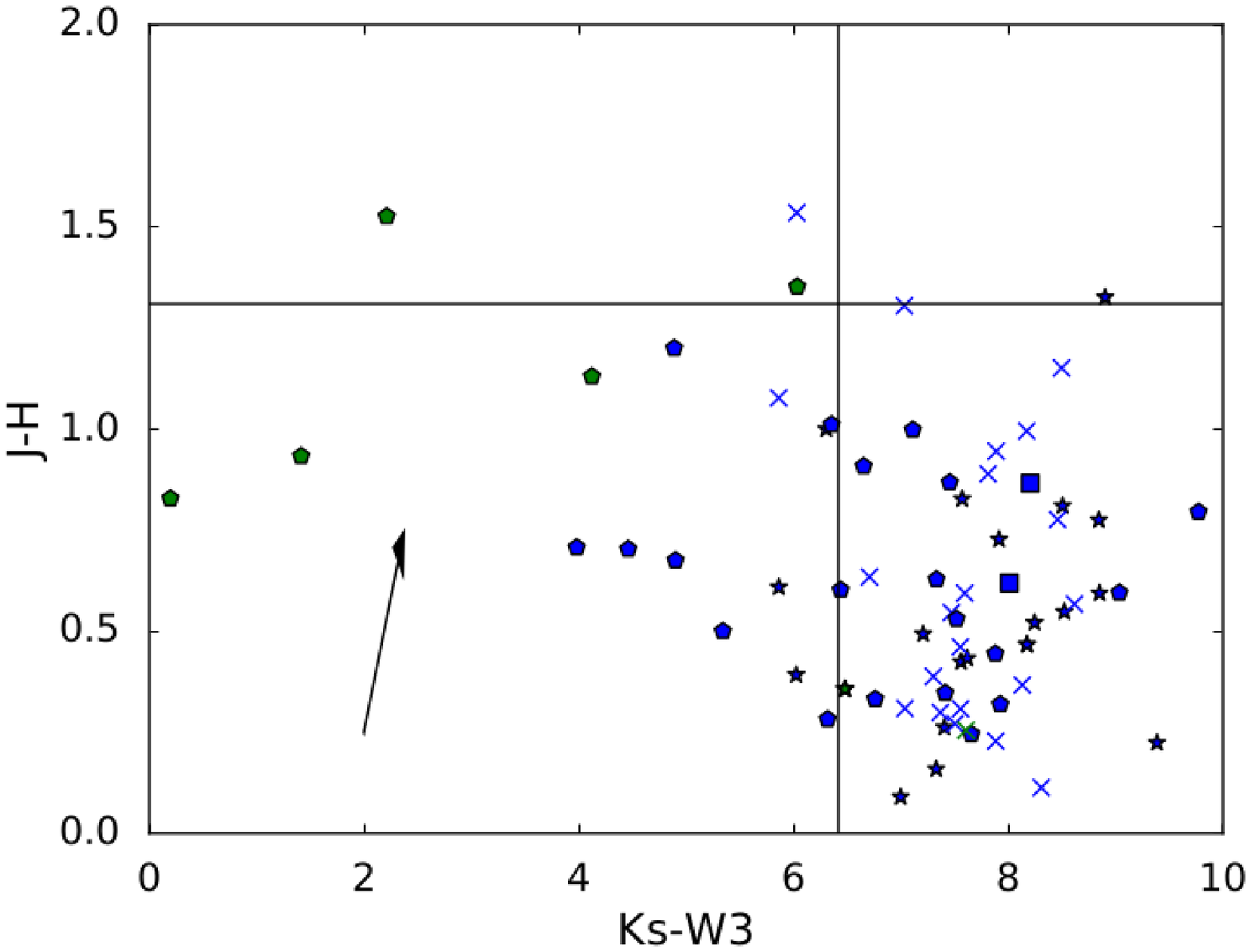}}
\caption[]{The distribution of HASH/IPHAS/VPHAS+ PNe with angular radius between 0 and 8~arcsec, for the colour indices criteria derived by the classification tree algorithm: M1 (upper-left panel); M2 (upper-right panel) and M3 (lower panel). The vertical and horizontal lines illustrate the criteria of each model (see Fig.~\ref{fig1}). Symbols are the same as in Figure~\ref{fig3}. The black arrows correspond to 4~mag extinction in the V band.}
\label{fig4}
\end{figure*}

The distribution of the HASH PNe is also explored in the {\it W1--W4}/{\it J--H}, {\it H--W2}/{\it J--H} and {\it Ks--W3}/{\it J--H} DCCDs, in Figure~\ref{fig4}. We find that the majority of them satisfies both criteria from M1 and M3. In particular, most of the PNe that does not satisfy these criteria have angular radius larger than 6~arcsec. Regarding the M2 model, all PNe with r$<$6~arcsec (except from two) have {\it H--W2}$\geq$2.24 and from them approximately half exhibit {\it J--H}$<$0.503. Sources with {\it J--H}$>$0.503 cannot be excluded from lists of PN candidates, but their contamination by YSOs and \hii~regions is high and thus it is very difficult to distinguish them without further information.

In addition to our classification tree criteria, we also restricted our final lists of candidate PNe by applying the following criteria: (i) r$<$19.5~mag; (ii) \ha, r$^\prime$ and i$^\prime$ measurements with errors lower than 0.1~mag, as proposed by the VPHAS+ group (or equivalent to signal-to-noise, S/N, $>$10) and (iii) 2MASS measurements with errors lower than 0.2~mag and AllWISE measurements with errors lower than 0.3~mag. Moreover, we visually inspected the AllWISE images of the candidates and we excluded those with uncertain WISE emission, specifically at 11.6 and 22.1~$\mu$m. Both emission bands are not always associated with a compact source, since they may also be associated with a more diffuse background emission. This significantly reduces the false identifications of YSOs as compact PNe.  

Overall, our classification tree models M1, M2 and M3 provide a list of compact PN candidates with high purity (less contamination from mimics) but low completeness (PNe with {\it J--H}$>$1.3 are not included in our list). 

\section{Discussion}
\subsection{Viironen's PNe candidate list}

\begin{figure*}
\vbox{
\includegraphics[width=\columnwidth]{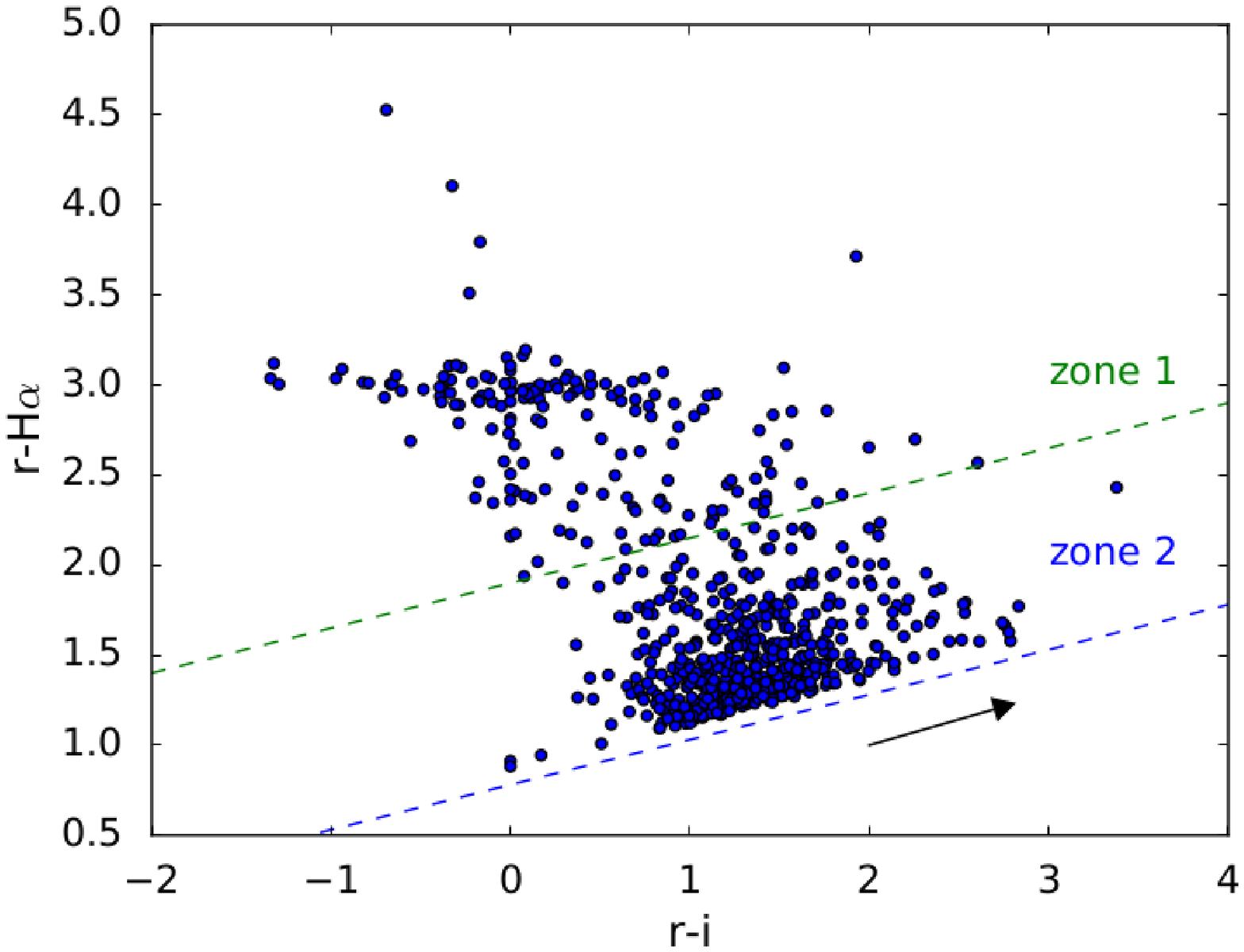}
\includegraphics[width=\columnwidth]{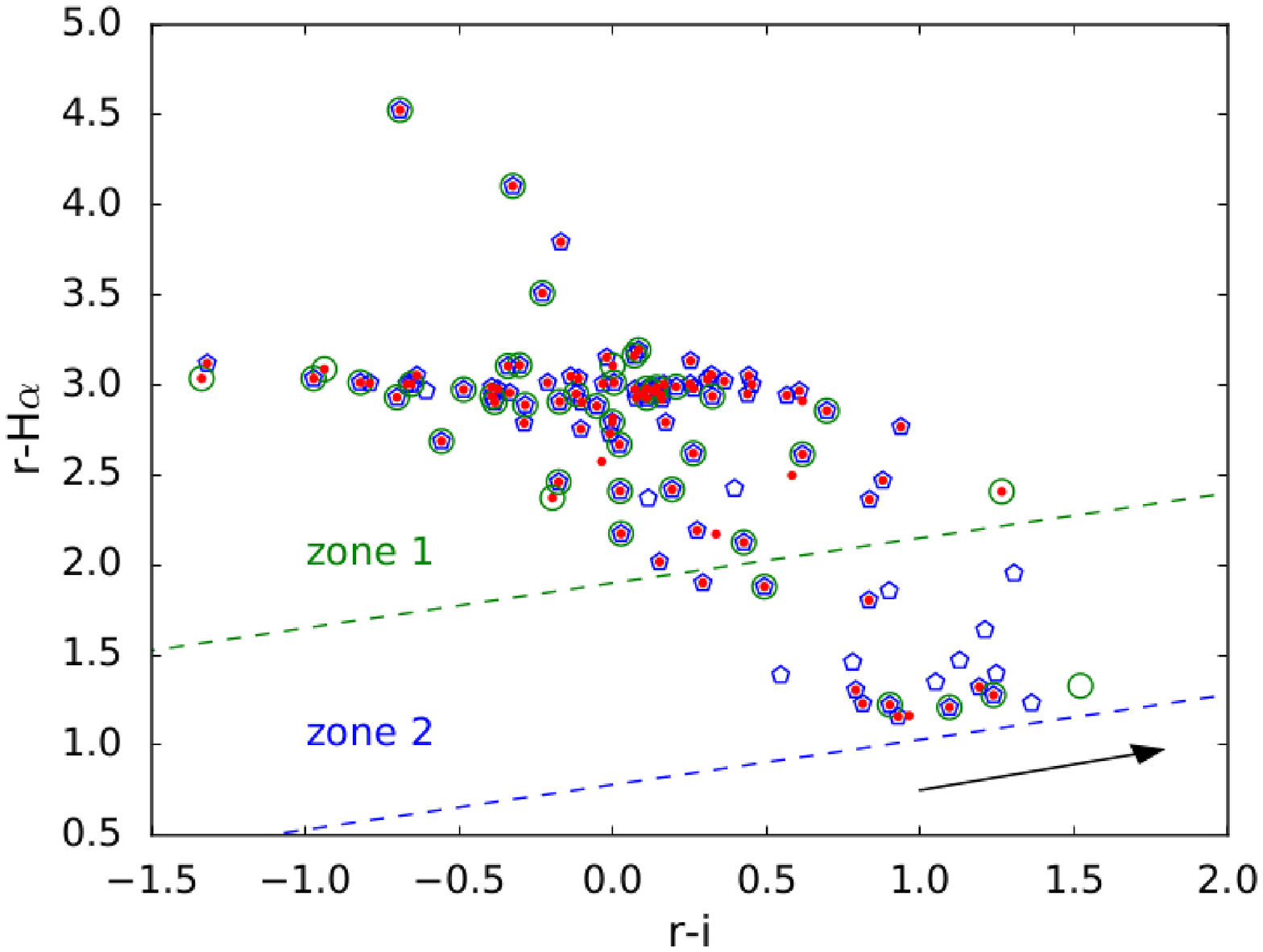}}
\caption[]{Right panel: r$^\prime$-\ha\ vs. r$^\prime$-i$^\prime$ DCCD for the catalogue of candidate IPHAS PNe \citep[]{Viironen2009b}. The green and blue dashed-lines delimit the zones 1 and 2 defined by \citet{Viironen2009b}. Left panel: r$^\prime$-\ha\ vs. r$^\prime$-i$^\prime$ DCCD only for the list of candidate PNe obtained by applying our three classification tree models. The symbols correspond to the three models (M1, red points; M2, green circles and M3, blue pentagons). The black arrows correspond to 3~mag extinction in the V band.}
\label{fig5}
\end{figure*}

First we applied our three classification tree models to the IPHAS list of PN candidates published by \citet{Viironen2009b}. The cross-matching with the 2MASS and AllWISE catalogues gave 671 matches with available 2MASS and AllWISE counterparts (letf panel, Figure \ref{fig5}). From this list, 91, 43 and 89 sources satisfy the criteria from the M1, M2 and M3 models, respectively. Their positions on the IPHAS DCCD is shown in the right panel of Figure \ref{fig5}.

\begin{figure*}
\vbox{
\includegraphics[width=\columnwidth]{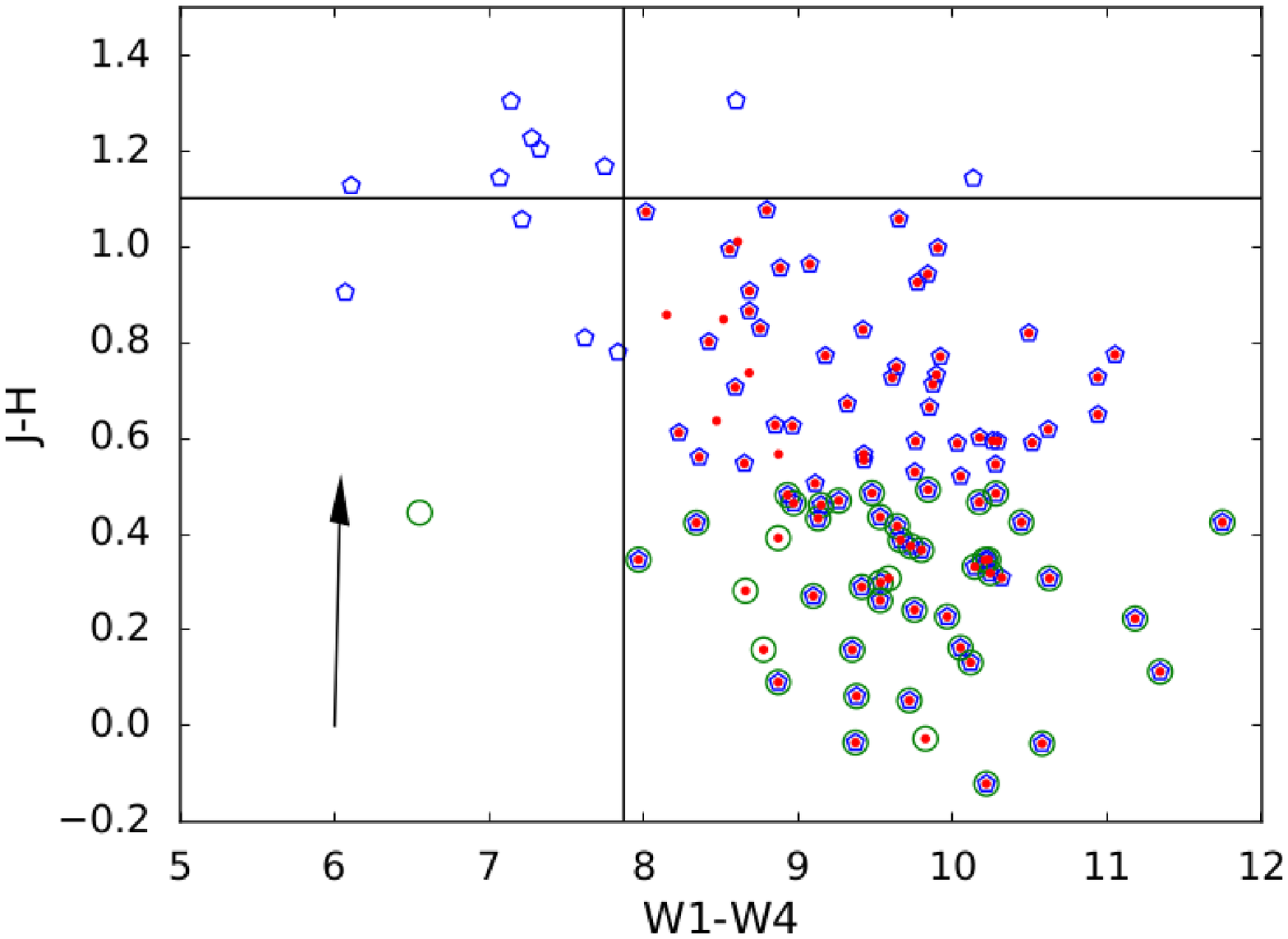}
\includegraphics[width=\columnwidth]{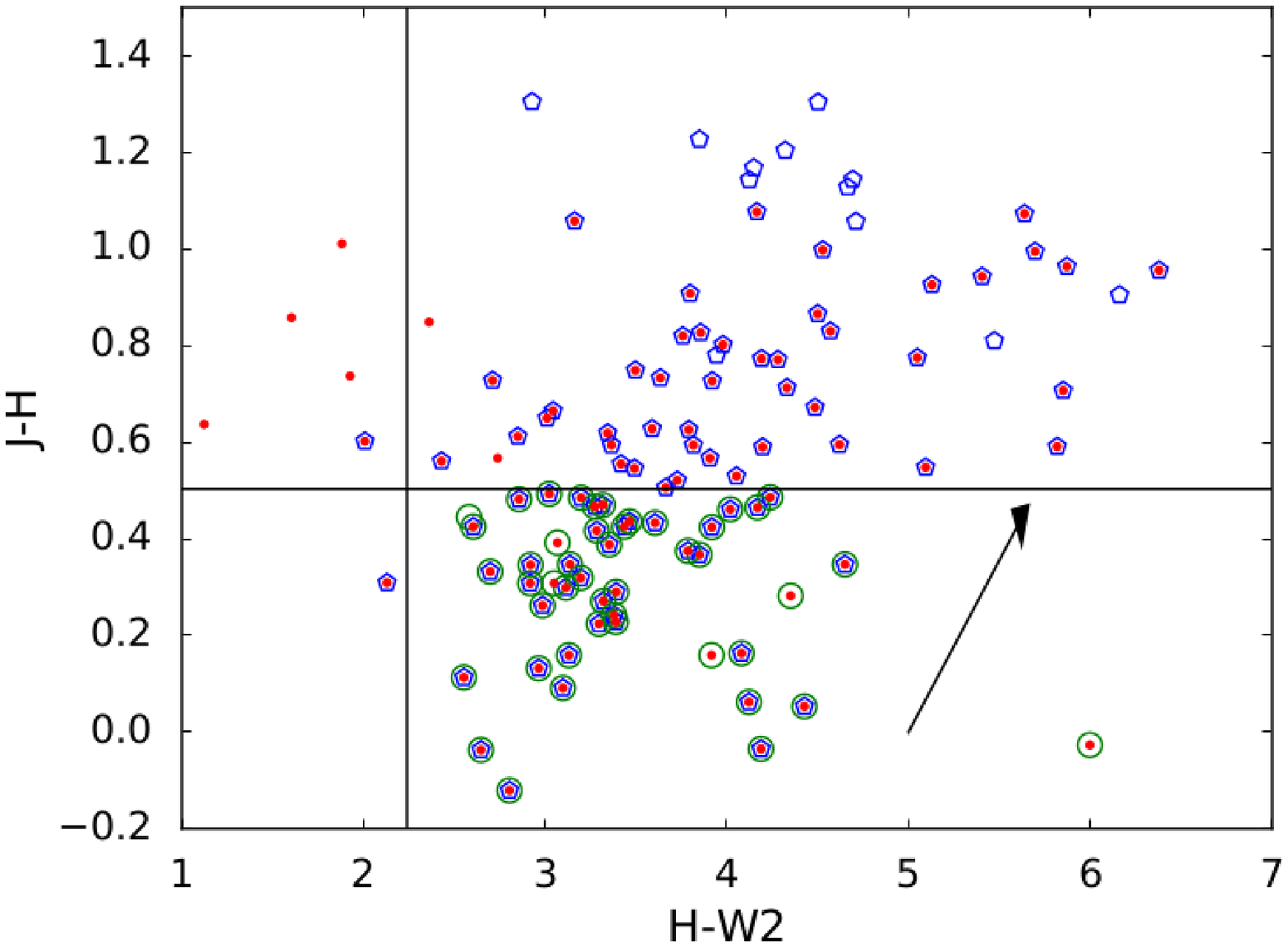}
\includegraphics[width=\columnwidth]{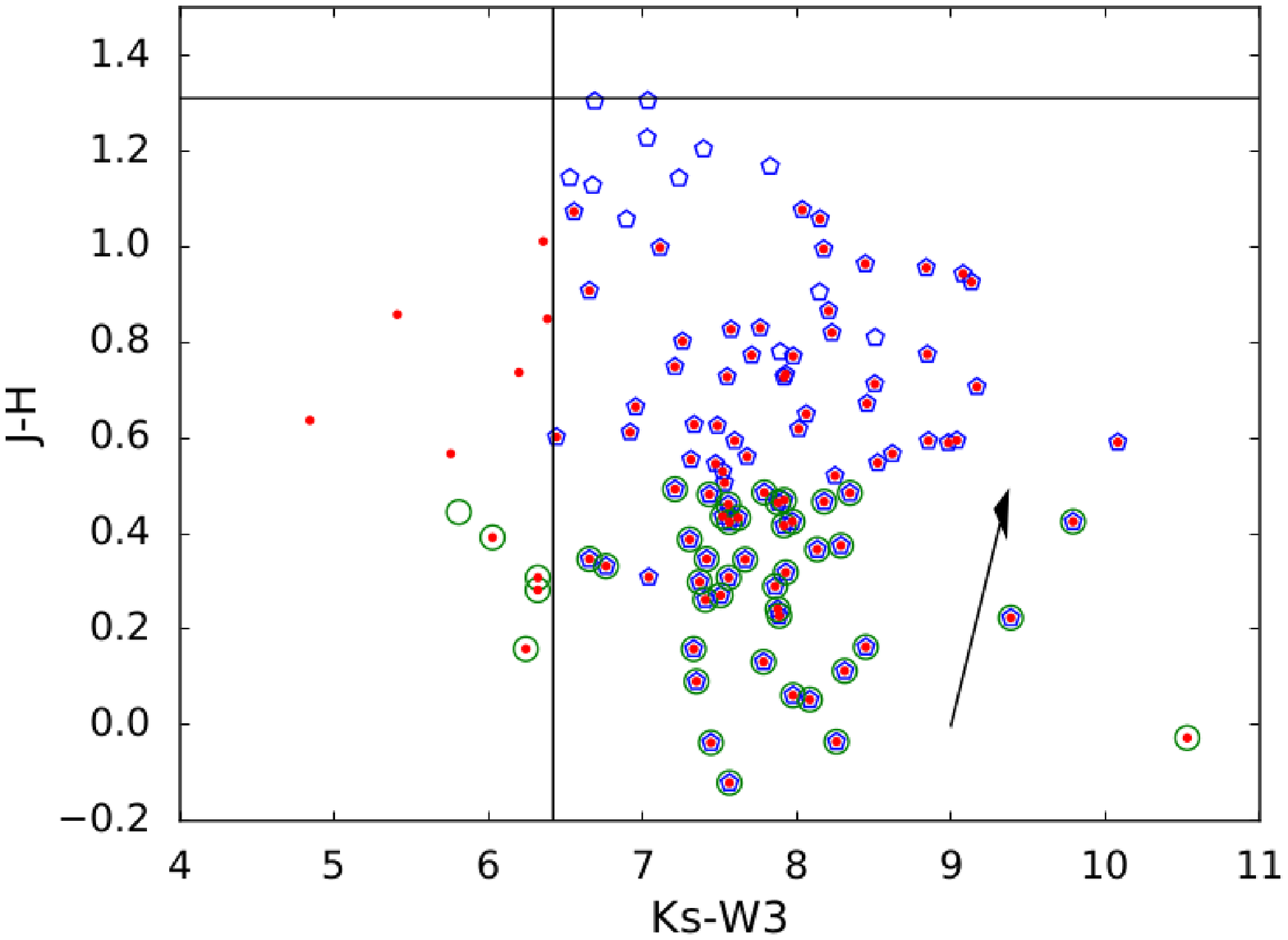}}
\caption[]{The distribution of the candidate IPHAS PNe \citep[]{Viironen2009b} for the colour indices criteria derived by the classification tree algorithm: M1 (upper left panel); M2 (upper right panel) and M3 (lower panel). The vertical and horizontal lines illustrate the criteria of each model (see Fig.1). The black arrows correspond to 4~mag extinction in the V band.}
\label{fig6}
\end{figure*}

The total number of sources that pass the criteria of at least one model is 99. Thirty-eight of the sources (38 per cent) satisfy the criteria of all the three models and are considered as candidates with high probability of being PNe, while 86 sources (87 per cent) satisfy the criteria from at least two models. Finally, 85 out of 99 sources are found to be strong \ha\ emitters occupying the zone~1 in the IPHAS DCCD. Figure \ref{fig6} displays the distribution of the sources recovered from the classification tree models in the IR DCCDs. In Table~\ref{table1}, we list all sources that satisfy at least one of our models, their coordinates, their classification in SIMBAD, the classification tree model they satisfy and the zone in the IPHAS DCCD they belong.

Based on their classification in SIMBAD, 81 out of 99 sources (82 per cent) turned out to be known PNe, while there are two classified as possible PNe, one as genuine and two as possible \hii~regions, two as emission line stars, three as infrared sources, one as AeBe star and seven are unclassified. Five of these unclassified sources are from the zone~1 indicative of strong \ha-emission sources. 

All the unclassified sources as well as those with an uncertain classification are so far considered as good PN candidates. Our technique wrongly classifies as PNe only two genuine and one possible \hii~region and one AeBe star. This implies low occurrence of false positive detection. 

Viironen's list is heavily contaminated from \ha~mimics. Around 50 per cent of the sources, for which follow-up spectroscopic data have been obtained, turned out to be emission line stars and they do not satisfy our criteria. Moreover, from the 34 spectroscopically confirmed IPHAS PNe in Viironen's catalogue, only 17 have 2MASS and AllWISE counterparts. We recovered 15 of them, or 88 per cent. Moreover, from the 149 known PNe in Viironen's list, our methodology recovers 81 of them while the remaining do not satisfy our criteria.

Seven IPHAS sources are also classified as SySts or possible SySts \citep{Viironen2009b}. Three of them are true SySts and they are recovered by the IR criteria of SySts \citep{Akras2019}. The remaining four sources do not satisfy the SySts's criteria and we consider them as non-symbiotic objects. For example, IPHASXJ184336.6+034640 is classified as a probable SySt but it satisfies the criteria of PN and  not those of SySts. It has a very high {\it W1-W4} colour index, which is indicative of a genuine PN.

Our list includes highly probable compact PNe and it is less contaminated from \ha-emitter mimics. However, the completeness of our list is low since it does not recover the entire population of compact PNe like those with high {\it J-H} colour index (see Figures \ref{fig2}, \ref{fig4}, and \ref{fig6}). 

\subsection{IPHAS DR2 catalogue}

Using the list of IPHAS PN candidates, all the sources are automatically restricted by satisfying the 2MASS criterion -- ({\it J--H})~$<$~1.64$\times$({\it H--Ks})-0.35 \citep[]{Viironen2009b}. According to our analysis through the classification tree algorithm, this 2MASS criterion should be substituted with our {\it J--H} criterion derived from our models. Viironen's 2MASS criterion provides sources with high {\it J--H} colour index as candidate PNe (see their Figure~1) and this is the main reason for the high contamination from mimics, as their follow-up spectroscopic observations have shown. This agrees with the statement of the authors that {\it \lq\lq ... there is a higher probability for our candidates to be PNe if they are located high up in the IPHAS two-colour diagram and towards low {\it J--H} and  intermediate {\it H--Ks} colours in the 2MASS two-colour diagram.\rq\rq} \citep[]{Viironen2009b}.

Therefore, we decided to apply our criteria to the entire IPHAS DR2 catalogue seeking for possible missing compact PNe. We ended up with 237 sources that satisfy the IPHAS and our classification tree criteria. This number was later reduced to 120 after the visually inspection of their AllWISE images. Only the objects that display a compact stellar source in all the AllWISE bands were considered as candidate PNe. The latter list includes all the aforementioned 99 sources (Table~\ref{table1}) and 21 new ones. 
 
The 21 additional sources found in the IPHAS DR2 catalogue are not included in Viironen's list (e.g. the known PN, RA:21 14 20.03~Dec.:+43 41 36.0) because of the violation of the Viironen's 2MASS criterion of saturation, their location at the borders of the CCDs and/or they were not detected at least twice. Our selection of PN candidates in the IPHAS DR2 catalogue is restricted only by the IPHAS criterion for \ha\ sources, whereas \citet{Viironen2009a,Viironen2009b} carried out a more detailed and robust selection. 
Nevertheless, in this short list of 21 IPHAS PN candidates, seven are known PNe, two are classified as possible PNe, one is classified as YSO, three as emission line stars and eight are unclassified. For most of them, the violation of Viironen's 2MASS criterion is the reason they were not included in the first list of PN candidates. In Table~\ref{table2}, we list the 21 sources from the IPHAS DR2 catalogue providing their coordinates, classification in SIMBAD, the classification three model they satisfy and the zone in the IPHAS DCCD they belong.

\subsection{VPHAS DR2 catalogue}

Besides IPHAS survey, its southern counterpart called VPHAS+ \citep[]{Drew2014} is still an on-going photometric survey. Our criteria are also applied to the VPHAS+ DR2 \citep[]{Drew2016} catalogue seeking for new PN candidates.

\begin{figure*}
\vbox{
\includegraphics[width=\columnwidth]{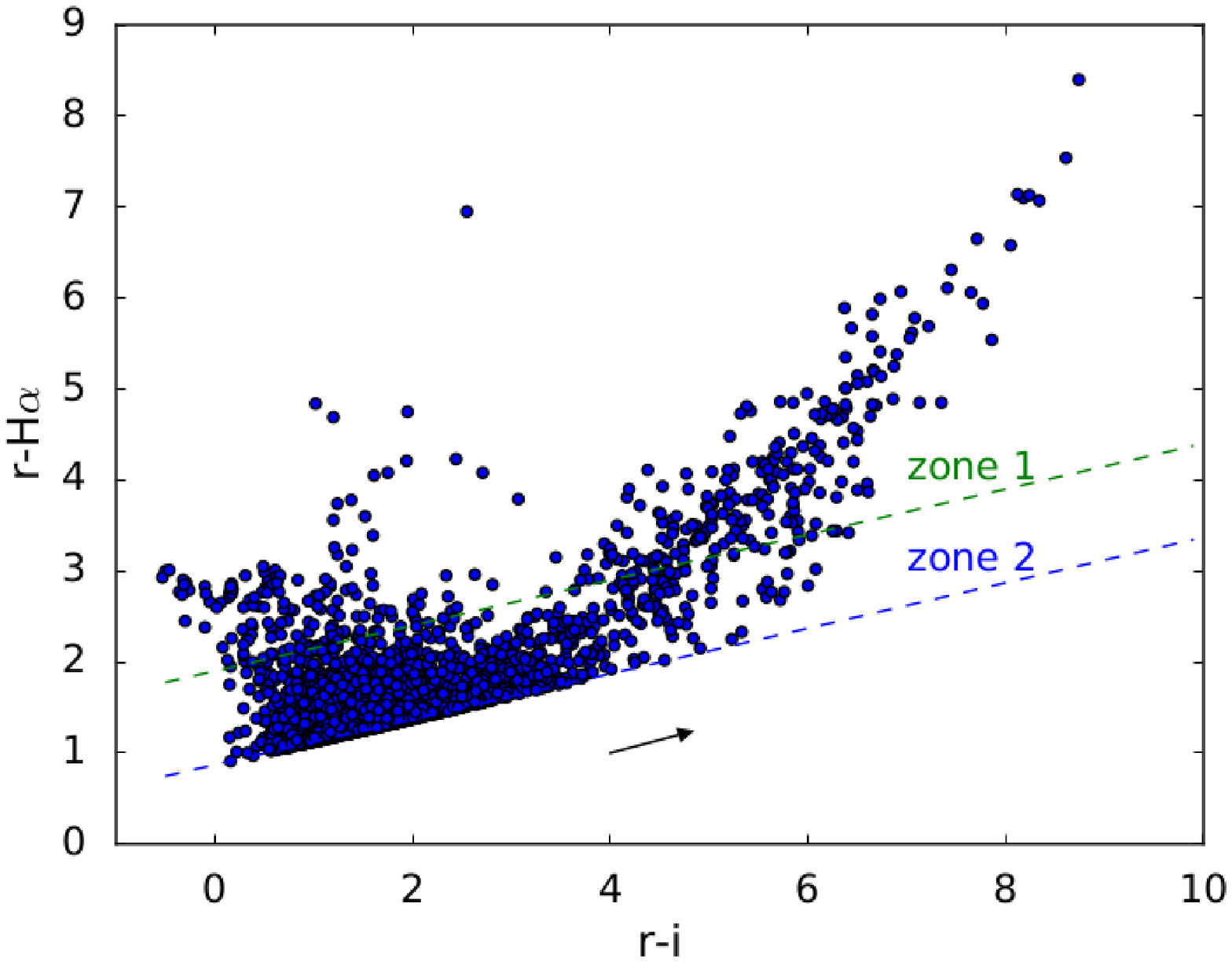}
\includegraphics[width=\columnwidth]{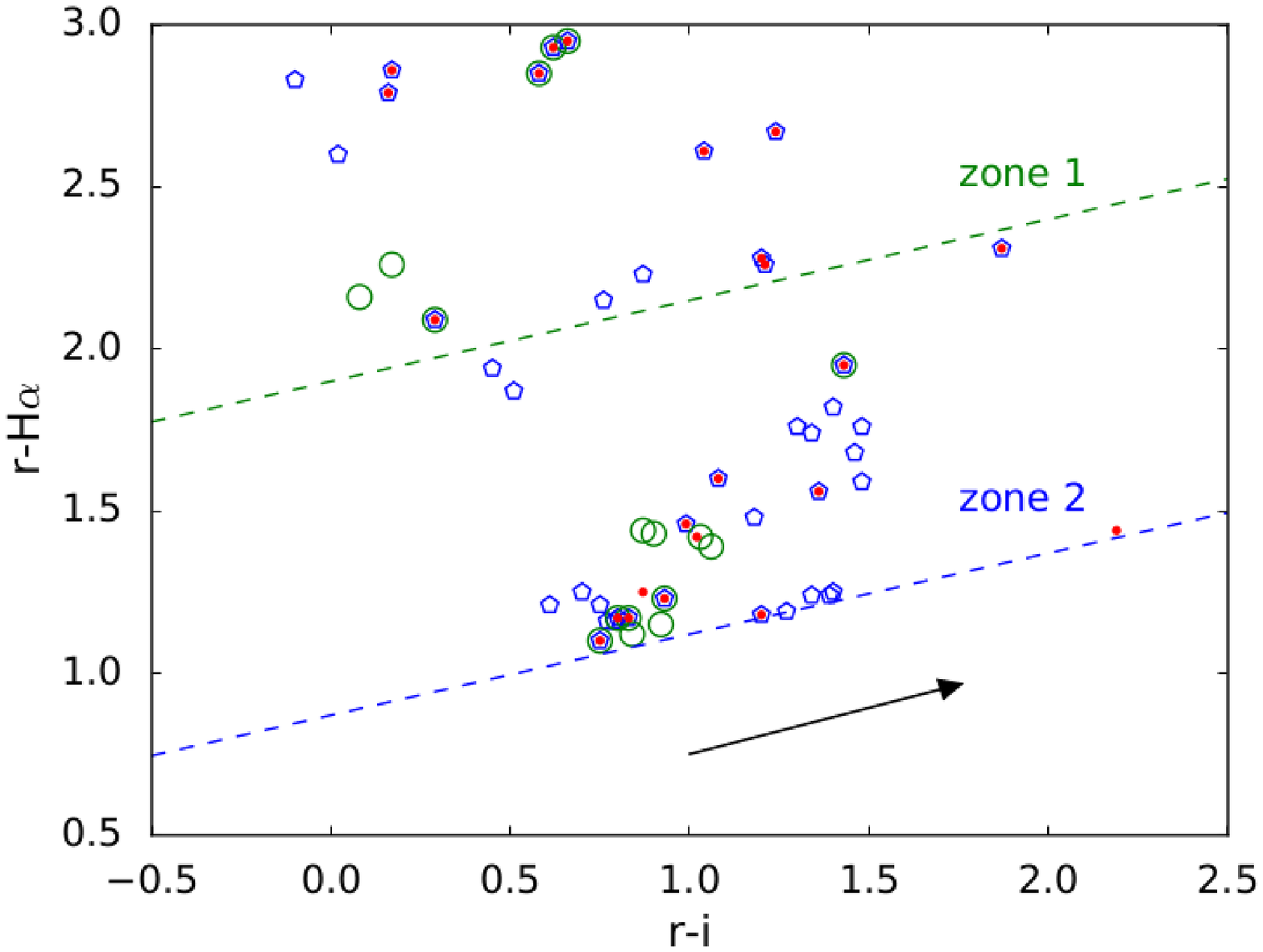}}
\caption[]{Right panel: r-\ha vs. r-i DCCD for the VPHAS+ DR2 catalogue. The green and 
blue dashed-lines delimit zones~1 and 2 defined by \citet{Viironen2009b}. Left panel: r-\ha vs. r-i DCCD only for the list of candidates derived after applying the criteria of our classification tree models. The different symbols correspond to the three models (M1, red points; M2, green circles; M3, blue pentagons). The black arrows correspond to 3~mag extinction in the V band.}
\label{fig7}
\end{figure*}

\begin{figure*}
\vbox{
\includegraphics[width=\columnwidth]{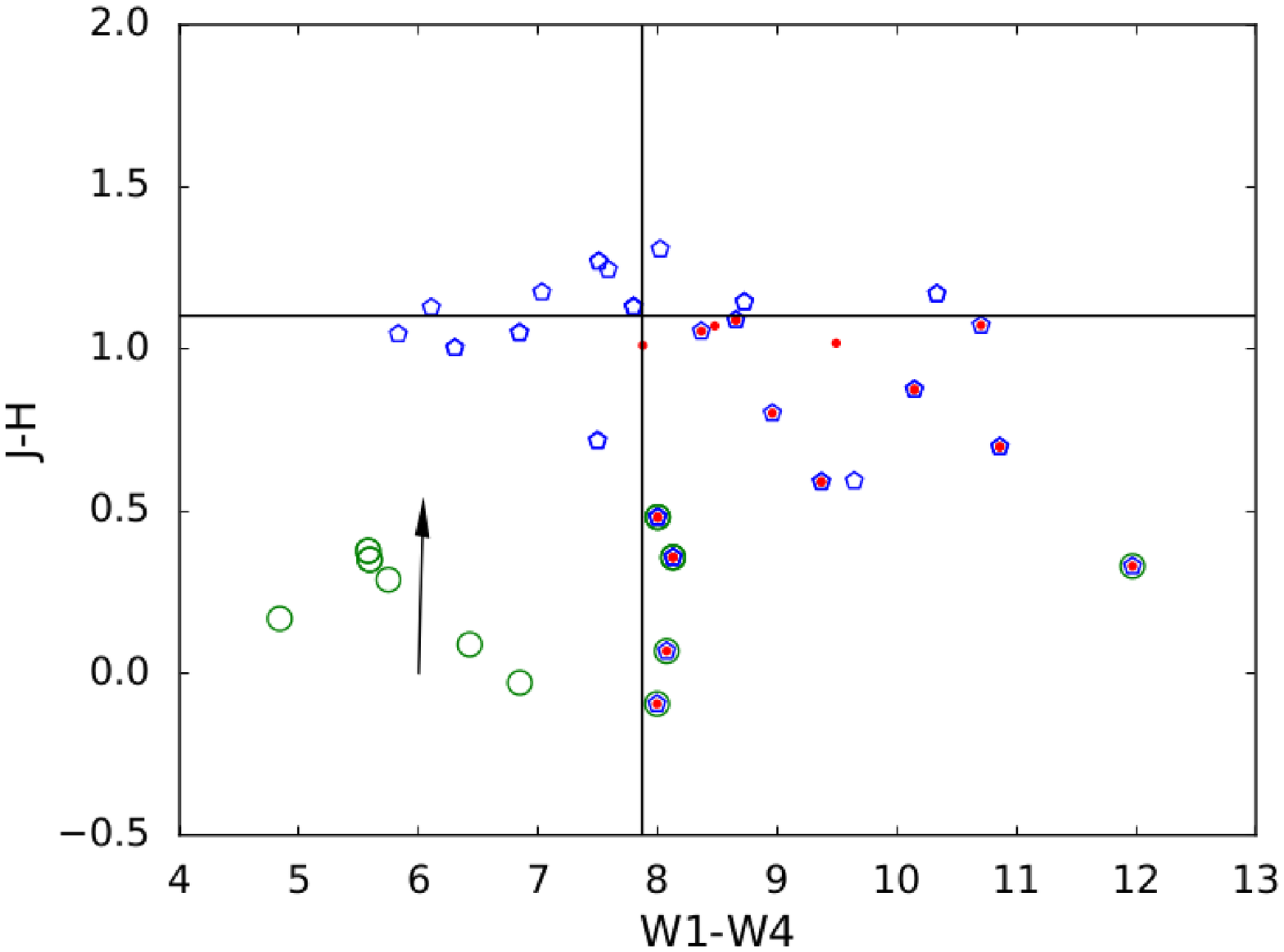}
\includegraphics[width=\columnwidth]{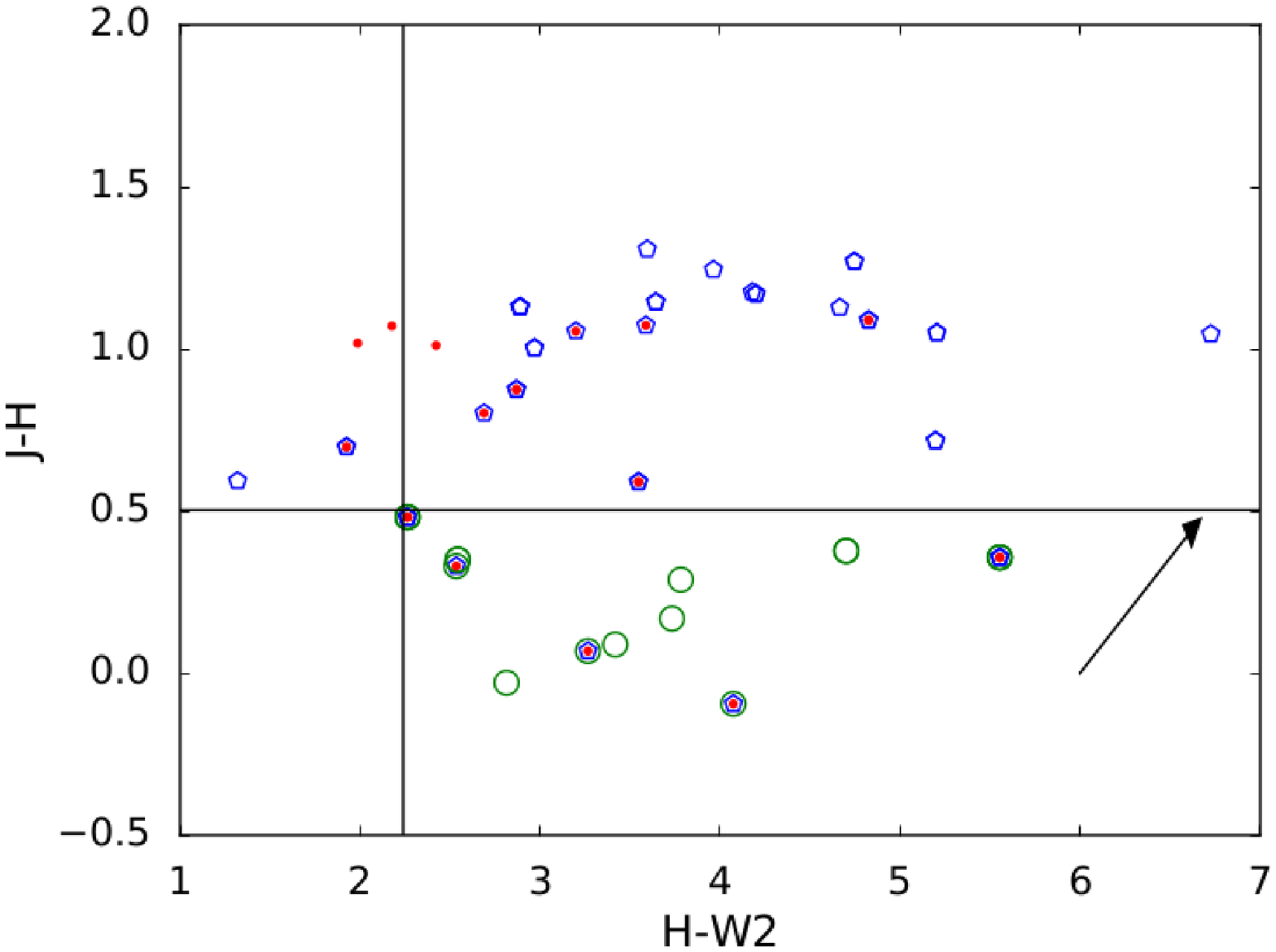}
\includegraphics[width=\columnwidth]{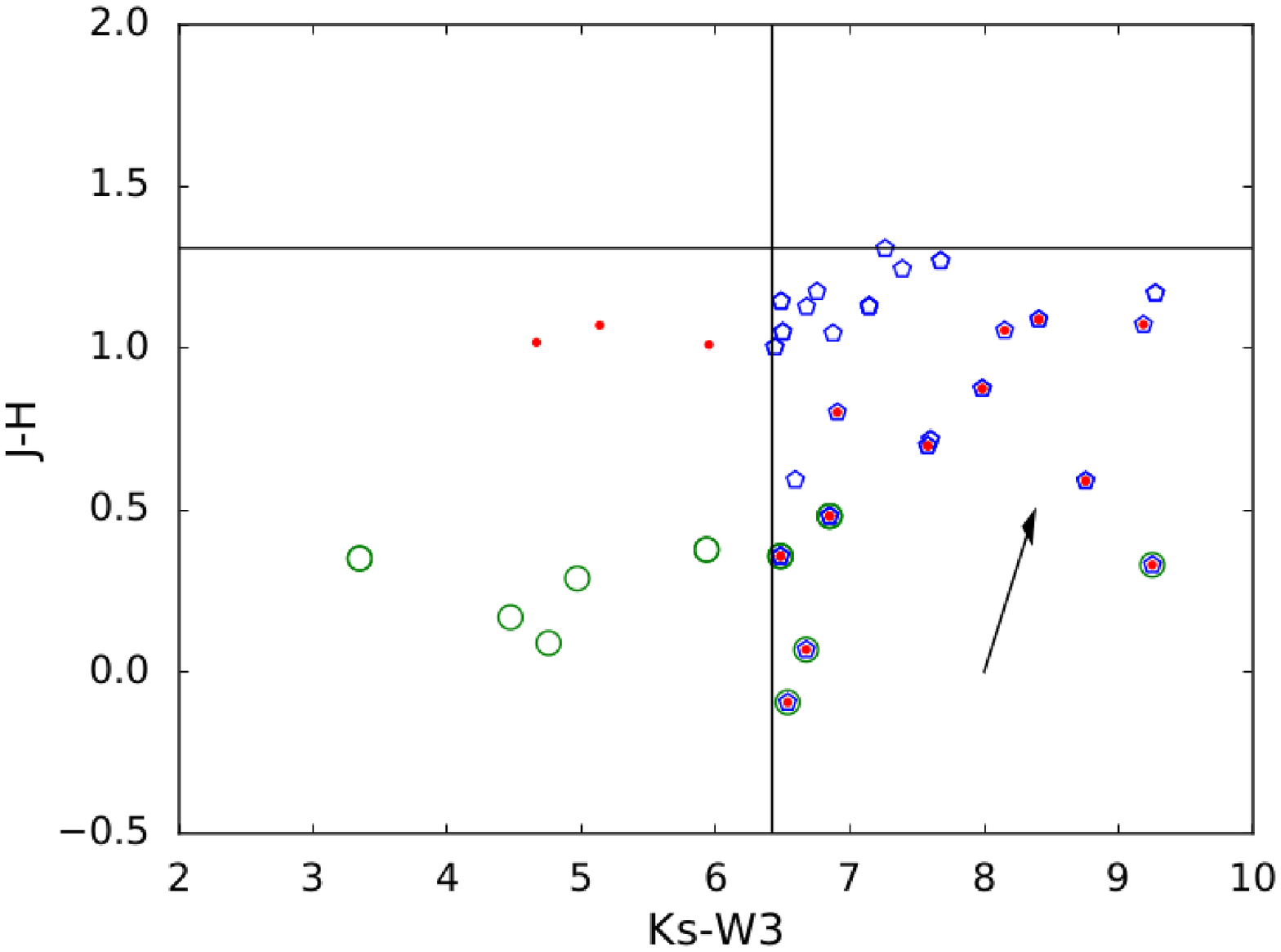}}
\caption[]{The distribution of the VPHAS+ candidate PNe for the colour indices criteria derived by the classification tree models: M1 (upper left panel); M2 (upper right panel) and M3 (lower panel).  The vertical and horizontal lines illustrate the criteria of each model (see Fig.1). The black arrows correspond to 4~mag extinction in the V band.}
\label{fig8}
\end{figure*}

Firstly, we cross-matched the VPHAS+ DR2 and AllWISE catalogues. We, secondly, applied the IPHAS criteria for the zones 1 and 2 in order to get only the \ha\ emitters (right panel, Figure~\ref{fig7}). Thus, thirdly, we applied our classification tree criteria to derive the final bulk of VPHAS+ candidate PNe. In the left panel of Figure~\ref{fig7}, we show the distribution of all the VPHAS+ PN candidates in the {\it r$^\prime$}-\ha\ vs. {\it r$^\prime$-i$^\prime$} DCCD. Their distributions in the  {\it W1--W4} vs. {\it J--H}, {\it H--W2} vs. {\it J--H} and {\it Ks--W3} vs. {\it J--H} DCCDs are presented in Figure~\ref{fig8}.  

Our first list of VPHAS+ PN candidates consisted of 34 sources. This number was reduced to 21 sources after a visual inspection of their AllWISE images. According to SIMBAD, these sources are classified as follow: 4 PNe, 3 possible PNe, 1 AGB star, 1 post-AGB star, 1 emission line star, 1 YSO  and 10 sources without classification (Table \ref{table3}). The contamination of our VPHAS+ list from other \ha-emitters appears to be low given that only one YSO was found. The remaining sources deserve a spectroscopic investigation in order to confirm their true nature.     

Sixteen of the candidates are found to occupy zone~1 (strong \ha\ emission) in the IPHAS DCCD and only five are from zone~2. For example, the post-AGB and AGB stars have {r$^\prime$--\ha} equal to 2.61 and 1.16~mag respectively. This implies that the post-AGB star is a strong \ha\ source (zone~1) and passes the criteria from two models (M1 and M3). On the other hand, the AGB star has a moderate \ha\ emission line (zone~2) and satisfies the criteria from only one model (M3). 

The VPHAS+ survey is still on-going and it's DR2 catalogue corresponds to only 24 per cent of the entire area, which corresponds to one-quarter of the observing area covered by IPHAS DR2. Assuming for simplicity that the stellar distribution in the Milky Way is the same in the northern and southern hemispheres (without taking into account the 200 additional square degrees of Galactic bulge in VPHAS+) and that both surveys, IPHAS and VPHAS, are almost identical -- r$^\prime$, i$^\prime$ and \ha\ magnitudes down to 20$^{th}$ magnitude, covered area $|${\it b}$|$ $<$5~degrees and arcsec spatial resolution -- an expected comparable population of PNe should to be found. For this first quartile of VPHAS+, 21 sources satisfy our criteria (if the above assumption holds, we expect at least 99 more sources after the end of the survey), while 120 sources are found in the IPHAS list.

In the simplest case that no additional compact PNe are found in the IPHAS, the VPHAS+ DR2 should include approximately 88 (excluding the confirmed non-PNe in IPHAS)/4=22 compact PNe. Given that 4 are already recovered, we expect 18 compact PNe in our VPHAS+ list. It should be noted that 13 sources are seemingly good PNe candidate: eight unclassified sources with strong \ha\ emission (zone~1), three possible PNe, one emission line star and one post-AGB star.

\subsection{LAMOST}

\begin{figure*}
\vbox{
\includegraphics[height=4.5cm, width=8cm]{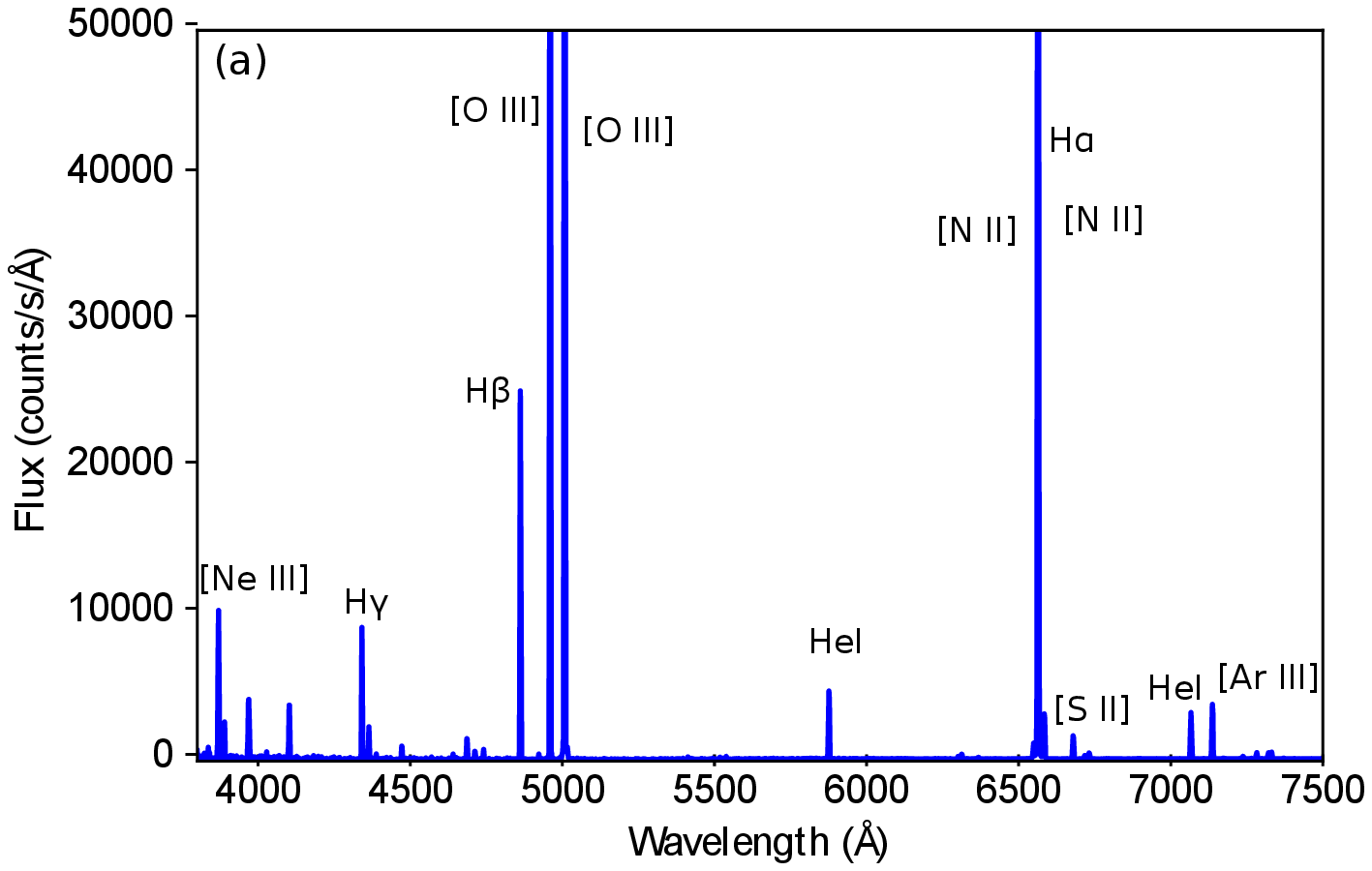}
\includegraphics[height=4.5cm, width=8cm]{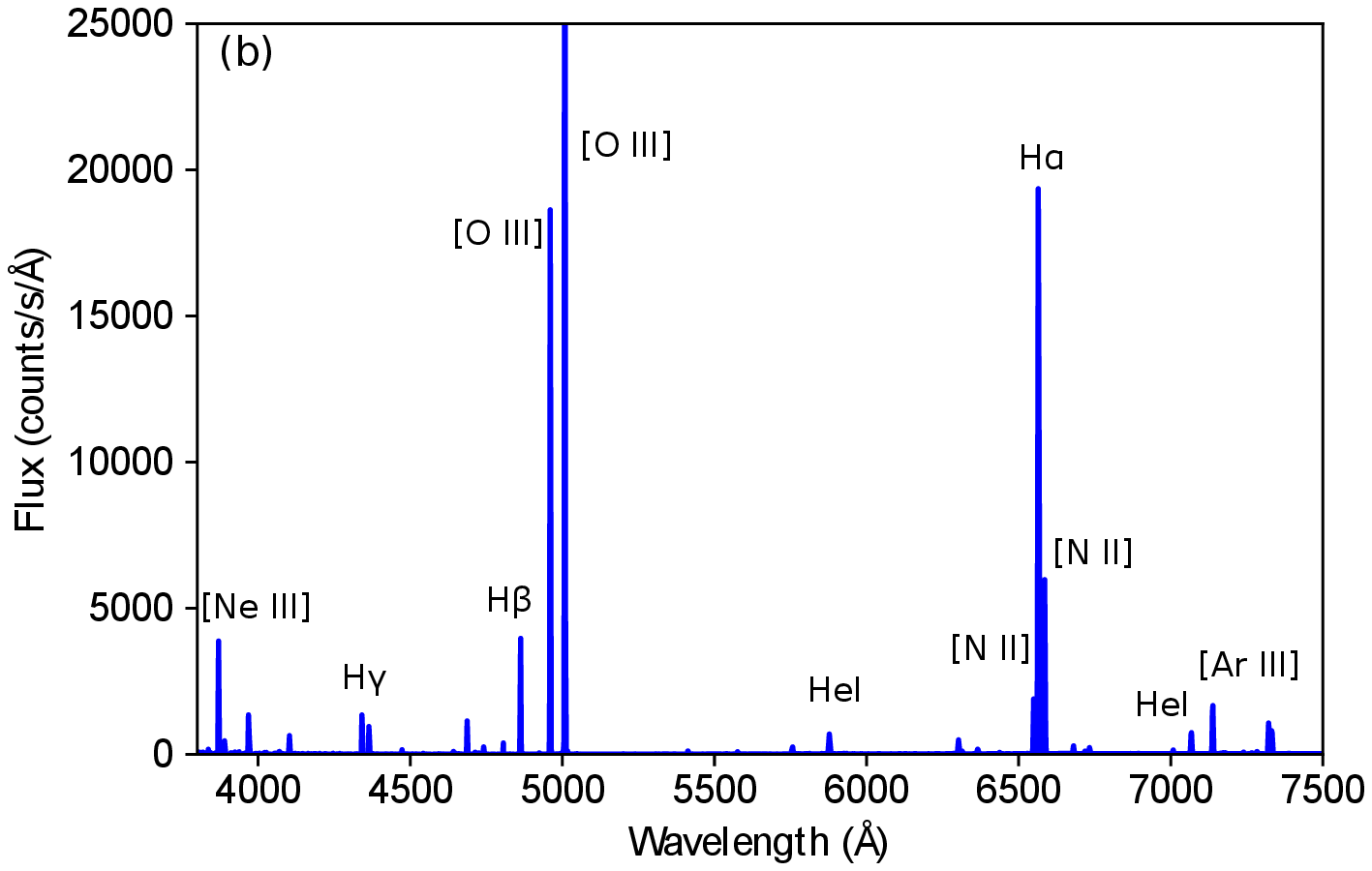}
\includegraphics[height=4.5cm, width=8cm]{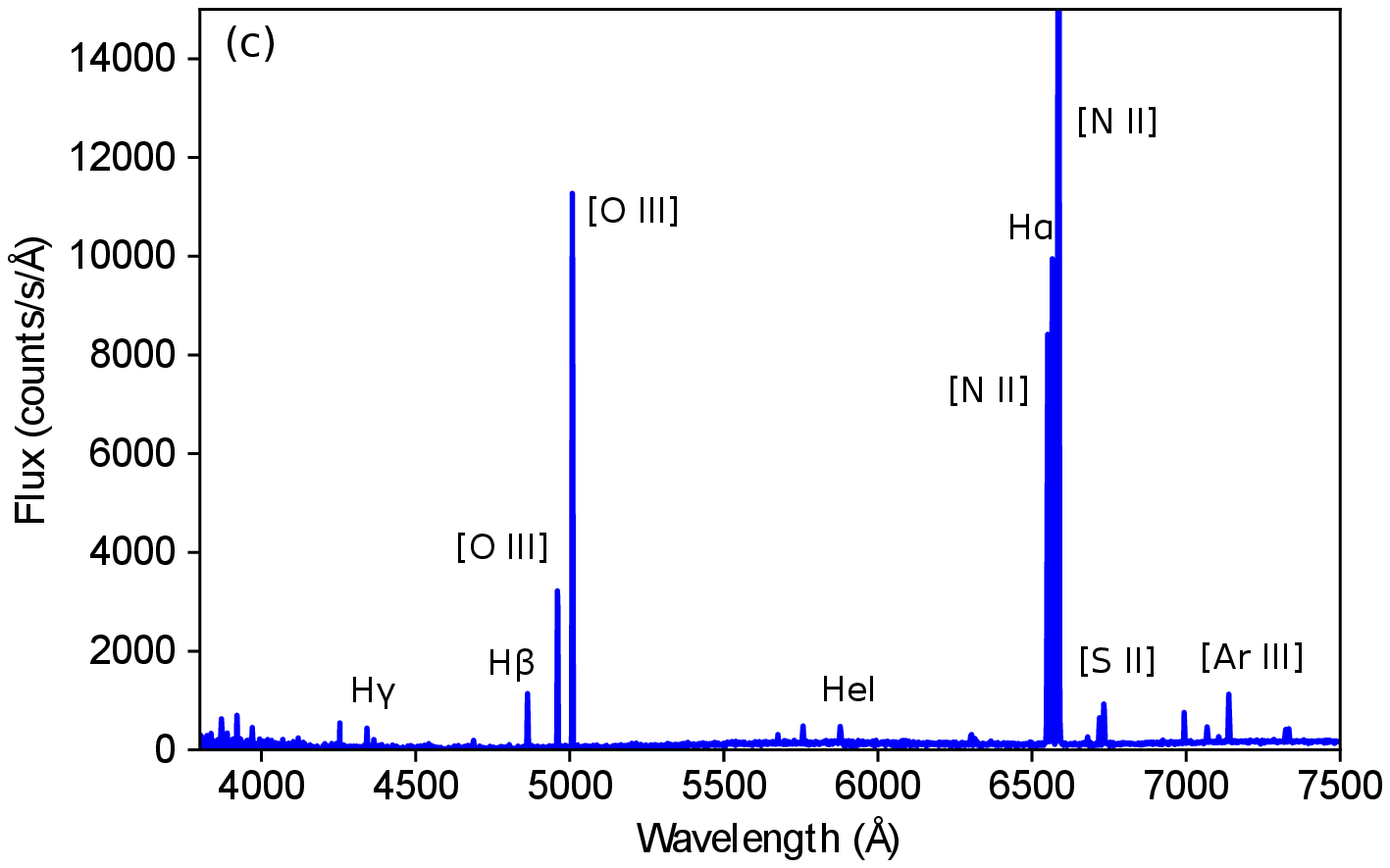}
\includegraphics[height=4.5cm, width=8cm]{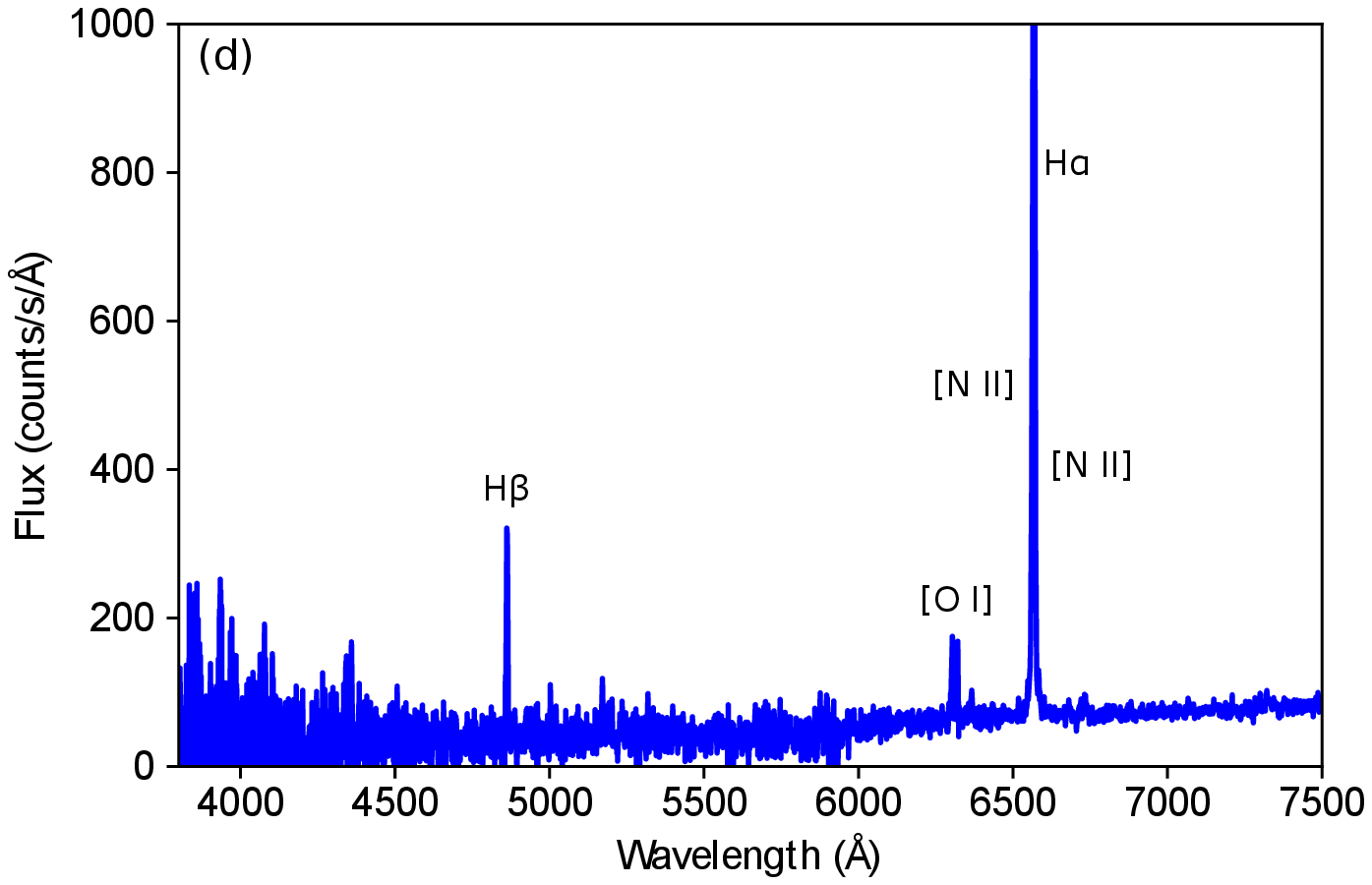}
\includegraphics[height=4.5cm, width=8cm]{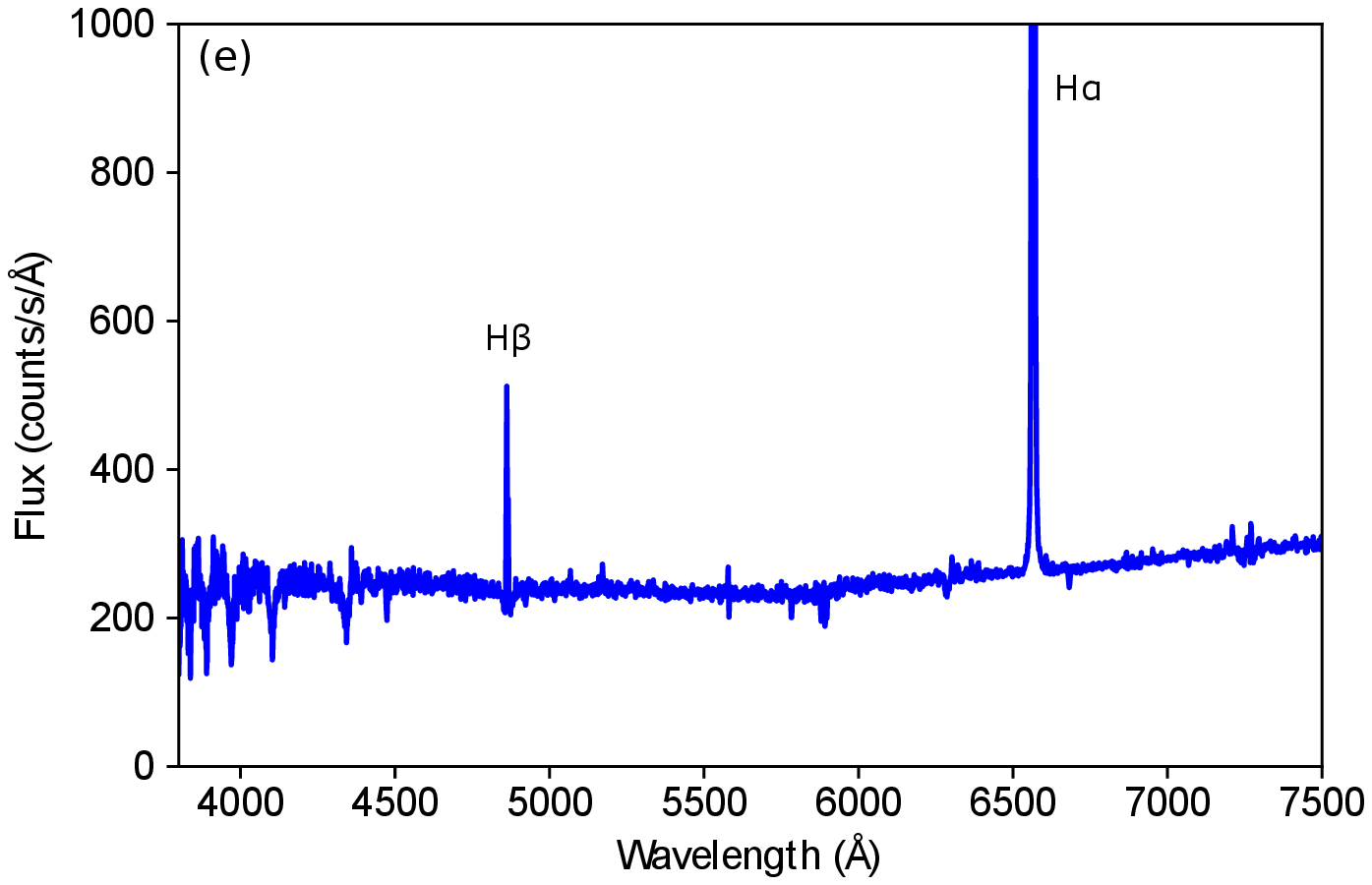}}
\caption[]{LAMOST DR4 spectra of the five IPHAS candidates, (a) RA:03 41 43, Dec:+52 17 00, (b) RA:05 41 22, Dec:+39 15 08, (c) RA:05 58 45, Dec:+25 18 43, (d) RA:06 07 11, Dec:+29 41 31, (e) RA:06 46 56, Dec:+01 16 40.}
\label{fig9}
\end{figure*}

The Large Sky Area Multi-Object Fiber Spectroscopic Telescope \citep[LAMOST; ][]{Zhao2012} is an innovative spectroscopic surveys with more than 9 millions of spectra from the northern hemisphere. By cross-matching our IPHAS lists of candidate PNe with the LAMOST DR4 catalogue (Luo et al. 2018), five matches were found. Three of them are known PNe, one is an Ae star, and one has no classification. Figure \ref{fig9} illustrates the LAMOST spectra of these five sources. The three PNe show, besides the Balmer lines, the \oxygeniii, \nitrogen and \sulfurt\ emission lines among others. The Ae star and not-classified candidate, on the other hand, display only lines from the Balmer series and low-excitation lines (e.g. \oxygeni~$\lambda$6300). This implies that the true nature of the unclassified candidate (RA:$\rm{06~07~11.19}$, Dec.:$\rm{+29~41~31.8}$) is more likely a YSO rather than a genuine compact PN. This is also supported by the low fluxes of the \ha\ line, which is consistent with the low {\it r$^\prime$-\ha} colour index and its location in the IPHAS zone~2. As mentioned above, the occurrence of finding genuine PNe in zone~2 is much lower than in zone~1.

\section{Conclusions}
A machine learning technique was devised to search for PN candidates in the IPHAS and VPHAS+ photometric catalogues. For this study, the 2MASS and AllWISE photometric data were combined with the classification tree algorithm in order to find suitable infrared criteria robust enough to distinguish PNe from other classes of \ha~emitters.

Three different classification tree models were found that identify PNe in the best possible way: {\it W1-W4}$\geq$7.87 and {\it J-H}$<$1.10 (M1); {\it H-W2}$\geq$2.24 and {\it J-H}$<$0.50 (M2) and {\it Ks-W3}$\geq$6.42 and {\it J-H}$<$1.31 (M3). 

Before applying these criteria to the IPHAS and VPHAS+ catalogues, the distribution of genuine PNe from the HASH catalogue in the IPHAS ({\it r$^\prime$}-\ha) vs. ({\it r$^\prime$-i$^\prime$}) DCCD was explored for different angular radii up to 8~arcsec. PNe with smaller angular sizes have higher {\it r$^\prime$}-\ha\ colour index and they belong to zone 1.

The application of our criteria to the list of IPHAS PNe candidates \citep{Viironen2009b} provided a list of 99 sources, from which 81 are genuine PNe (82~per cent) and 11 are likely compact PNe. Repeating the same technique for the entire IPHAS DR2 catalogue, 21 new candidates, not included in Viironen's list, were recovered and seven of them turned out to be known PNe (33 per cent) and 13 likely compact PNe. Finally, 21 VPHAS+ sources were also found to pass our criteria and only four of them are known PNe, while 15 of them are likely compact PNe.

Overall, 39 (24 in the IPHAS and 15 in the VPHAS+) sources of uncertain or without classification in SIMBAD were found to satisfy our IR criteria as well as IPHAS criteria and they are considered as good PN candidates. The follow-up spectroscopic analysis of these sources will be presented in a forthcoming paper. Finally, more than 80 compact PNe are expected to be found in the final VPHAS+ catalogue after the end of the survey.

\section*{Acknowledgements}
The authors would like to thank the referee, Dr. Walter Weidmann, for his constructive comments and suggestions. LGR is supported by NWO funding towards the Allegro group at Leiden University. DRG thanks the partial support of CNPq (grant 304184/2016-0). This paper makes use of data from the Two Micron All-Sky Survey which is a joint project of the University of Massachusetts and the Infrared Processing and Analysis Center/California Institute of Technology, funded by the NASA and the National Science Foundation, data products from the Wide-field Infrared Survey Explorer, which is a joint project of the University of California, Los Angeles, and the Jet Propulsion Laboratory/California Institute of Technology, funded by the National Aeronautics and Space Administration. It also makes use of data obtained as part of the INT Photometric \ha\ Survey of the Northern Galactic Plane (IPHAS, www.iphas.org) carried out at the Isaac Newton Telescope (INT). The INT is operated on the island of La Palma by the Isaac Newton Group in the Spanish Observatorio del Roque de los Muchachos of the Instituto de Astrof\'isica de Canarias. All IPHAS data are processed by the Cambridge Astronomical Survey Unit, at the Institute of Astronomy in Cambridge. The band merged DR2 catalogue was assembled at the Centre for Astrophysics Research, University of Hertfordshire, supported by STFC grant ST/J001333/1. Based on data products from observations made with ESO Telescopes at the La Silla Paranal Observatory under programme ID 177.D-3023, as part of the VST Photometric \ha Survey of the Southern Galactic Plane and Bulge (VPHAS+, www.vphas.eu). Information from the Red MSX Source survey database at website which was constructed with support from the Science and Technology Facilities Council of the UK were also used. Guoshoujing Telescope (the Large Sky Area Multi-Object Fiber Spectroscopic Telescope LAMOST) is a National Major Scientific Project built by the Chinese Academy of Sciences. Funding for the project has been provided by the National Development and Reform Commission. LAMOST is operated and managed by the National Astronomical Observatories, Chinese Academy of Sciences. This research has also made use of the SIMBAD database, operated at CDS, Strasbourg, France. Finally, the following software packages in Python were used: Matplotlib\citep{Hunter2007}, NumPy \citep{Walt2011}, SciPy \citep{jones2001} and AstroPy Python \citep{Astropy2013,Astropy2018,Muna2016}.




\bibliographystyle{mnras}
\bibliography{references} 



\appendix

\section{Results}

\onecolumn
\begin{longtable}{lllllll} 
\caption{List of known and candidates PNe that satisfy our IR criteria obtained 
from \citet{Viironen2009b}.} \\
\label{table1}
RA (J2000.0) &   Dec (J2000.0) & SIMBAD class. &  {\it W1--W4} & {\it H--W2} & {\it Ks--W3}  & Comments\\
\hline
\endhead
\hline
\endfoot
\hline
00 05 53.45 &  64 05 15.6  &  IR      &  \xmark & \xmark & \cmark  & zone 2\\
00 20 17.37 &  59 18 39.8  & IR/p H~II$^{*}$ &  \cmark & \xmark & \cmark  & zone 1\\
01 45 51.23 &  64 16 05.7  &  p PN    &  \xmark & \xmark & \cmark  & zone 2\\ 
03 41 43.40 &  52 17 00.4  &  PN      &  \cmark & \cmark & \cmark  & zone 1\\
04 13 15.00 &  56 56 58.3  &  PN      &  \cmark & \xmark & \cmark  & zone 2\\
04 15 54.54 &  48 49 40.5  &  PN      &  \xmark & \xmark & \cmark  & zone 1\\
04 20 45.26 &  56 18 12.6  &  PN      &  \cmark & \xmark & \cmark  & zone 1\\
05 10 44.09 &  47 10 04.5  &  ?       &  \cmark & \xmark & \cmark  & zone 2\\
05 41 22.11 &  39 15 08.1  &  PN      &  \cmark & \xmark & \cmark  & zone 1\\
05 46 50.01 &  24 22 02.8  &  PN      &  \cmark & \cmark & \cmark  & zone 1\\
05 52 42.85 &  26 21 16.2  &  PN      &  \cmark & \cmark & \cmark  & zone 2\\
05 58 45.32 &  25 18 43.9  &  PN      &  \cmark & \cmark & \xmark  & zone 1\\
06 17 19.67 &  24 51 25.0  &  ?       &  \xmark & \xmark & \cmark  & zone 2\\
06 35 45.13 & -00 05 37.3  &  PN      &  \cmark & \cmark & \cmark  & zone 1\\
06 38 16.13 & -01 33 08.3  &  IR      &  \cmark & \xmark & \cmark  & zone 2\\
06 38 21.34 &  09 00 32.7  &  ELS     &  \xmark & \xmark & \cmark  & zone 2\\
06 39 55.87 &  11 06 30.8  &  PN      &  \cmark & \cmark & \cmark  & zone 1\\
06 46 56.42 &  01 16 40.8  &  AeBe    &  \cmark & \xmark & \xmark  & zone 2\\
18 32 49.69 & -00 56 38.4  &  PN      &  \cmark & \cmark & \cmark  & zone 1\\
18 33 17.48 &  00 11 46.7  &  PN      &  \cmark & \xmark & \cmark  & zone 1\\
18 34 38.70 &  00 08 02.8  &  PN      &  \cmark & \cmark & \cmark  & zone 1\\
18 41 40.44 & -01 25 18.1  &  PN      &  \cmark & \cmark & \cmark  & zone 1\\
18 43 36.60 &  03 46 40.2  &  PN/p SySt$^{\dag}$ &  \cmark & \xmark & \cmark  & zone 1\\
18 48 52.99 &  01 28 52.2  &  PN      &  \cmark & \xmark & \cmark  & zone 1\\
18 50 05.69 & -00 40 41.2  &  p PN/p SySt$^{\dag}$& \xmark & \xmark & \cmark  & zone 2\\
18 51 41.55 &  09 54 52.4  &  PN      &  \cmark & \xmark & \cmark  & zone 1\\
18 52 25.55 & -00 33 26.4  &  PN      &  \cmark & \xmark & \cmark  & zone 1\\
18 53 09.43 &  07 52 40.4  &  PN      &  \cmark & \cmark & \cmark  & zone 1\\
18 54 48.30 & -01 39 21.9  &  PN      &  \cmark & \xmark & \xmark  & zone 1 \\
18 56 07.34 &  13 31 33.2  &  ?       &  \cmark & \xmark & \xmark  & zone 1\\
18 56 18.18 &  07 07 26.2  &  PN      &  \cmark & \xmark & \cmark  & zone 1\\
19 00 34.81 & -02 11 58.0  &  PN      &  \cmark & \xmark & \cmark  & zone 1\\ 
19 01 36.05 &  00 00 10.5  &  PN      &  \cmark & \xmark & \cmark  & zone 1 \\
19 01 36.60 & -01 19 08.0  &  PN      &  \cmark & \cmark & \cmark  & zone 1\\
19 02 10.16 & -01 48 45.4  &  PN      &  \cmark & \xmark & \xmark  & zone 1\\
19 02 37.10 & -00 26 57.2  &  PN      &  \cmark & \cmark & \cmark  & zone 1\\
19 11 35.83 &  13 31 11.3  &  PN      &  \cmark & \xmark & \cmark  & zone 1\\
19 12 05.82 &  15 09 04.3  &  PN      &  \cmark & \cmark & \cmark  & zone 1\\
19 13 34.55 &  04 38 03.8  &  PN      &  \cmark & \cmark & \cmark  & zone 1\\
19 14 04.14 &  17 31 32.8  &  PN      &  \cmark & \cmark & \cmark  & zone 1\\
19 16 21.44 &  16 56 38.5  &  PN      &  \cmark & \xmark & \cmark  & zone 1\\
19 16 27.70 &  05 13 19.0  &  PN      &  \cmark & \cmark & \cmark  & zone 1\\
19 17 27.27 &  14 27 34.6  &  PN      &  \cmark & \xmark & \cmark  & zone 1\\
19 17 50.55 &  08 15 08.4  &  PN      &  \cmark & \cmark & \cmark  & zone 1\\
19 19 02.65 &  19 02 20.8  &  PN      &  \cmark & \cmark & \cmark  & zone 1\\
19 19 42.91 &  16 21 27.8  &  PN      &  \cmark & \xmark & \cmark  & zone 1\\
19 22 26.66 &  10 41 21.3  &  PN      &  \cmark & \xmark & \cmark  & zone 1\\
19 23 24.81 &  21 08 00.1  &  PN      &  \cmark & \cmark & \cmark  & zone 1\\
19 23 46.86 &  21 06 38.1  &  PN      &  \cmark & \cmark & \cmark  & zone 1\\
19 24 22.21 &  09 53 56.2  &  PN      &  \cmark & \cmark & \cmark  & zone 1\\
19 26 37.75 &  21 09 27.0  &  PN      &  \cmark & \xmark & \cmark  & zone 1\\
19 27 44.02 &  21 30 03.4  &  PN      &  \cmark & \xmark & \cmark  & zone 1\\
19 27 44.81 &  10 24 20.6  &  PN      &  \cmark & \cmark & \cmark  & zone 1\\
19 29 02.62 &  24 46 47.0  &  PN      &  \cmark & \cmark & \xmark  & zone 1\\
19 30 16.64 &  14 47 21.6  &  PN      &  \cmark & \xmark & \cmark  & zone 1\\
19 31 16.48 &  10 03 21.4  &  PN      &  \cmark & \cmark & \cmark  & zone 1\\
19 33 09.04 &  22 58 33.5  &  PN      &  \cmark & \xmark & \cmark  & zone 1\\
19 33 46.75 &  24 32 26.8  &  PN      &  \cmark & \cmark & \cmark  & zone 1\\
19 38 52.08 &  25 05 33.4  &  PN      &  \xmark & \xmark & \cmark  & zone 2\\
19 39 15.95 &  16 20 48.0  &  PN      &  \cmark & \xmark & \cmark  & zone 1\\
19 39 35.81 &  20 19 02.0  &  PN      &  \cmark & \cmark & \cmark  & zone 1\\
19 41 09.29 &  14 56 58.9  &  PN      &  \cmark & \cmark & \cmark  & zone 1\\
19 41 33.97 &  17 45 17.5  &  PN      &  \cmark & \cmark & \cmark  & zone 1\\
19 45 22.15 &  21 20 03.8  &  PN      &  \cmark & \xmark & \cmark  & zone 1\\
19 45 32.91 &  23 28 09.9  &  PN      &  \cmark & \xmark & \cmark  & zone 1 \\
19 48 15.01 &  28 07 28.6  &  p H~II  &  \cmark & \xmark & \cmark  & zone 1\\
19 48 26.40 &  22 08 37.0  &  PN      &  \cmark & \cmark & \cmark  & zone 1\\
19 49 57.59 &  23 26 00.2  &  ?       &  \cmark & \cmark & \cmark  & zone 1\\
19 50 28.51 &  25 54 28.9  &  PN      &  \cmark & \xmark & \cmark  & zone 1\\
19 51 52.73 &  32 59 17.8  &  PN      &  \cmark & \cmark & \cmark  & zone 1\\
19 52 48.83 &  25 53 59.2  &  PN      &  \cmark & \xmark & \cmark  & zone 1 \\
19 55 49.80 &  31 13 39.4  &  ?       &  \cmark & \xmark & \xmark  & zone 1\\
19 59 12.66 &  33 50 03.6  &  ?       &  \xmark & \xmark & \cmark  & zone 1\\
20 00 42.07 &  32 27 40.8  &  PN      &  \cmark & \xmark & \cmark  & zone 1\\
20 03 11.44 &  30 32 33.9  &  PN      &  \cmark & \xmark & \cmark  & zone 1\\
20 03 22.45 &  27 00 54.8  &  PN      &  \cmark & \xmark & \cmark  & zone 1\\
20 04 44.21 &  31 27 26.7  &  PN      &  \cmark & \xmark & \cmark  & zone 1\\
20 04 58.63 &  25 26 37.1  &  PN      &  \cmark & \xmark & \cmark  & zone 1\\
20 10 37.67 &  31 37 56.7  &  ?       &  \cmark & \xmark & \cmark  & zone 1\\
20 12 24.03 &  40 45 29.2  &  PN      &  \cmark & \cmark & \cmark  & zone 1\\
20 12 33.68 &  40 47 40.5  &  H~II    &  \cmark & \xmark & \cmark  & zone 2\\
20 12 47.69 &  34 20 32.2  &  PN      &  \cmark & \cmark & \cmark  & zone 1\\
20 13 57.88 &  29 33 56.0  &  PN      &  \cmark & \xmark & \cmark  & zone 1 \\
20 15 22.20 &  40 34 44.6  &  PN      &  \cmark & \xmark & \cmark  & zone 1\\
20 19 54.23 &  43 05 59.6  &  PN      &  \cmark & \xmark & \cmark  & zone 1 \\
20 21 03.76 &  32 29 24.1  &  PN      &  \cmark & \xmark & \cmark  & zone 1\\
20 25 04.87 &  33 34 50.2  &  PN      &  \cmark & \cmark & \cmark  & zone 1\\
20 45 22.73 &  50 22 39.6  &  PN      &  \cmark & \cmark & \cmark  & zone 1\\
21 07 39.67 &  40 57 52.1  &  PN      &  \cmark & \cmark & \cmark  & zone 2\\
21 19 37.20 &  54 53 28.8  &  ELS     &  \cmark & \xmark & \cmark  & zone 2\\
21 20 44.88 &  51 53 27.4  &  H~II    &  \cmark & \cmark & \xmark  & zone 1\\
21 30 00.76 &  54 27 27.2  &  PN      &  \cmark & \cmark & \cmark  & zone 1\\
21 31 50.18 &  52 33 51.6  &  PN      &  \cmark & \cmark & \cmark  & zone 1\\
21 35 43.85 &  50 54 17.0  &  PN      &  \cmark & \cmark & \xmark  & zone 1\\
21 43 17.61 &  50 25 14.4  &  PN      &  \cmark & \cmark & \cmark  & zone 1\\
21 57 41.81 &  51 41 39.0  &  PN      &  \cmark & \cmark & \cmark  & zone 1\\
22 20 16.62 &  58 14 16.6  &  PN      &  \cmark & \xmark & \cmark  & zone 1\\
22 55 06.95 &  56 42 31.0  &  PN      &  \cmark & \xmark & \cmark  & zone 1\\
22 56 19.79 &  57 09 21.0  &  PN      &  \cmark & \cmark & \xmark  & zone 1\\
\end{longtable}
\medskip{}
\begin{flushleft}
$^{*}$ no classification is provided for this sources by \citet{Viironen2009b}.\\
$^{\dag}$ not a SySt according to the criteria from \cite{Akras2019}\\
p PN: possible PN, ELS: emission line star, SySt: Symbiotic star\\
\end{flushleft} 

\begin{longtable}{lllllll} 
\caption{List of known and candidates PNe that satisfy our IR criteria obtained  from the IPHAS DR2 catalogue}\\
\label{table2}
RA (J2000.0) &   Dec (J2000.0) & SIMBAD class. &  {\it W1--W4} & {\it H--W2} & {\it Ks--W3}  & Comments\\    			
\hline 
\endhead
\hline
\endfoot
\hline
 02 41 35.93  & +57 37 38.0   &  p PN/p SySt$^{\dag}$      &  \xmark & \xmark & \cmark  & zone 2, (a)\\  
 02 46 26.04  & +60 06 17.8   &  ?      &  \xmark & \xmark & \cmark  & zone 2, (a) \\   
 04 40 27.17  & +50 28 29.5   &  YSO      &  \cmark & \xmark & \cmark  & zone 2, (a)\\ 
 06 07 11.19  & +29 41 31.8   &  ?      &  \xmark & \xmark & \cmark  & zone 2, (a)\\   
 06 32 08.44  & +04 53 09.5   &  ELS      &  \cmark & \xmark & \xmark  & zone 1, (a)\\
 07 05 19.20  & +02 46 59.4   &  PN      &  \cmark & \cmark & \cmark  & zone 1, (b)\\   
 19 04 38.63  & +02 14 24.1   &  ELS      &  \cmark & \xmark & \xmark  & zone 2, (a)\\ 
 19 13 05.43  & +15 46 40.0   &  ?      &  \cmark & \cmark & \cmark  & zone 2, (c)\\   
 19 13 22.60  & +03 25 00.1   &  PN      &  \cmark & \xmark & \xmark  & zone 2, (a)\\   
 19 13 38.42  & +14 59 19.1   &  PN      &  \cmark & \cmark & \cmark  & zone 1, (b)\\   
 19 14 59.71  & +17 22 46.1   &  PN      &  \xmark & \xmark & \cmark  & zone 2, (b)\\   
 19 31 03.43  & +26 59 47.0   &  ?      &  \xmark & \xmark & \cmark  & zone 2, (a)\\   
 19 34 24.64  & +22 33 50.8   &  ?      &  \cmark & \xmark & \xmark  & zone 2, (a)\\   
 19 36 06.38  & +23 42 46.8   &  ?      &  \xmark & \xmark & \cmark  & zone 2, (a)\\   
 19 47 51.90  & +31 18 18.2   &  PN      &  \cmark & \cmark & \cmark  & zone 2, (a)\\   
 20 00 52.91  & +34 28 22.1   &  p PN      &  \cmark & \xmark & \cmark  & zone 2, (a)\\ 
 20 09 29.29  & +33 02 27.8   &  ?      &  \cmark & \xmark & \cmark  & zone 2, (c) \\   
 20 13 39.04  & +33 15 07.0   &  PN      &  \cmark & \xmark & \xmark  & zone 2, (a)\\   
 21 14 20.03  & +43 41 36.0   &  PN      &  \cmark & \xmark & \cmark  & zone 2, (a)\\   
 21 36 46.48  & +56 27 16.5   &  ELS      &  \cmark & \cmark & \cmark  & zone 2, (a)\\ 
 21 49 38.22  & +56 54 36.7   &  ?      &  \cmark & \xmark & \cmark  & zone 2, (c)\\ 
\end{longtable}
\medskip{}
\begin{flushleft}
$^{\dag}$  It is included in the IPHAS list of SySt candidates \citep[]{Corradi2008} but it does not satisfy the criteria of being a SySt proposed by \citet{Akras2019}.\\
(a) The 2MASS criterion used by \citet{Viironen2009b} is violated.\\
(b) The 2MASS criterion used by \citet{Viironen2009b} is not violated but \ha, r$^\prime$ and/or i$^\prime$ is saturated.\\
(c) The 2MASS criterion used by \citet{Viironen2009b} is not violated, \ha, r$^\prime$ and/or i$^\prime$ are not saturated, hence these sources should be included in the list of Viironen's, except they are located at the borders of the CCDs or they were not detected 
at least twice. 
\end{flushleft} 

\begin{longtable}{lllllll} 
\caption{List of known and candidates PNe that satisfy our IR criteria obtained from the VPHAS+ DR2 catalogue}\\
\label{table3}
\endhead
\hline
\endfoot
RA (J2000.0) &   Dec (J2000.0) & SIMBAD class. &  {\it W1--W4} & {\it H--W2} & {\it Ks--W3}  & Comments\\    			
\hline
07 54 26.38  &  -28 37 44.5    &  ?      &  \xmark & \cmark & \xmark  & zone 1\\ 
07 54 03.46  &  -26 47 29.6    &  ?      &  \xmark & \cmark & \xmark  & zone 1\\ 
10 04 40.05  &  -56 08 37.1    &  PN     &  \xmark & \xmark & \cmark  & zone 2\\ 
10 42 48.17  &  -59 25 28.9    &  YSO    &  \xmark & \xmark & \cmark  & zone 2\\    
10 53 51.09  &  -61 10 32.8    &  ?      &  \cmark & \xmark & \cmark  & zone 1\\ 
11 03 59.68  &  -61 03 27.7    &  ?      &  \cmark & \xmark & \cmark  & zone 1\\ 
11 09 25.74  &  -61 04 09.6    &  ?      &  \xmark & \xmark & \cmark  & zone 1\\ 
14 11 46.27  &  -64 16 23.9    &  PN     &  \cmark & \cmark & \cmark  & zone 1\\ 
16 19 40.18  &  -49 13 59.0 &  WR/p PN$^\dag$&  \cmark & \xmark & \xmark  & zone 1\\ 
16 38 01.79  &  -49 27 18.9 &  p PN/p SySt$^{\dag\dag}$& \cmark & \xmark & \cmark  & zone 1\\   
16 53 30.94  &  -41 09 44.2     &  ?      &  \xmark & \cmark & \xmark  & zone 2\\    
17 27 47.13  &  -28 11 03.3     &  ?      &  \xmark & \xmark & \cmark  & zone 2\\ 
17 32 22.57  &  -28 14 28.9     &  ?      &  \xmark & \xmark & \cmark  & zone 1\\ 
17 56 24.32  &  -29 38 06.7     &  PN     &  \xmark & \cmark & \xmark  & zone 1\\ 
18 02 05.24  &  -25 15 33.7     &  ELS    &  \cmark & \xmark & \cmark  & zone 1\\ 
18 03 31.20  &  -27 48 27.0     &  PN     &  \xmark & \xmark & \cmark  & zone 1\\ 
18 05 09.14  &  -19 28 11.2     &  p-AGB  &  \cmark & \xmark & \cmark  & zone 1\\ 
18 16 39.97  &  -17 04 35.7     &  AGB    &  \xmark & \xmark & \cmark  & zone 1\\ 
18 19 53.90  &  -11 48 45.8     &  ?      &  \cmark & \xmark & \cmark  & zone 1\\ 
18 46 16.33  &  -03 06 26.0     &  ?      &  \cmark & \cmark & \cmark  & zone 1\\ 
18 50 05.70  &  -00 40 41.1     &  p PN   &  \xmark & \xmark & \cmark  & zone 2\\ 
\hline
\end{longtable}
\medskip{}
\begin{flushleft}
$^\dag$ not PN according to \cite{Frew2014}\\
$^{\dag\dag}$ not a SySt according to the criteria from \cite{Akras2019}\\
\end{flushleft} 
\bsp	
\label{lastpage}
\end{document}